 \documentclass[manuscript]{aastex}

\slugcomment{published ApJ, 2010}

\shorttitle{Tracing the Galactic Environment of the Sun with ENAs}
\shortauthors{Frisch et al.}

\def\counts{counts cm$^{-2}$ sec$^{-1}$ sr$^{-1}$ keV$^{-1}$}
\def\BdotR{${ \bf B} \cdot {\bf  R} \sim 0$}
\def\Fmodtwo{$F_\mathrm{Mod2}$}
\def\Fmodthree{$F_\mathrm{Mod3}$}
\def\Fmodfour{$F_\mathrm{Mod4}$}
\def\abs{$|$}
\def\Pmag{$P_\mathrm{mag}$}
\def\Pthr{$P_\mathrm{ther}$}
\def\nHImodtwo{$n$(H$^\circ$)$_\mathrm{Mod2}$}
\def\nHImodthree{$n$(H$^\circ$)$_\mathrm{Mod3}$}
\def\nHImodfour{$n$(H$^\circ$)$_\mathrm{Mod4}$}
\def\dFesa3{$dF_\mathrm{1 \sigma}$}
\def\delmod{$|\Delta F_\mathrm{model}|$}
\def\vel{$V$}
\def\B{${B}$}
 
\def\microG{$\mu$G}

\def\glong{$\ell$}
\def\glat{$b$}
\def\elong{$\lambda$}
\def\elat{$\beta$}

\def\HI{H$^{\rm o}$}
\def\FeII{Fe$^{\rm +}$}
\def\MgII{Mg$^{\rm +}$}

\def\HeI{He$^{\rm o}$}

\def\nHI{$n \mathrm{(H^\circ)}$}

\def\CaII{$ \mathrm{N(Ca^+})$}

\def\npro{$n \mathrm{(p^+)}$}
\def\temp{$T$}

\def\kms{\hbox{km s$^{-1}$}}
\def\deeg{\hbox{$^{\rm o}$}}

\def\HeI{He$^{\rm o}$}
\def\cmtwo{cm$^{-2}$}
\def\cc{cm$^{-3}$}
\begin{document}

\title{Can IBEX Identify Variations in the Galactic Environment of the Sun using Energetic Neutral Atom (ENAs)? }

\author{Priscilla C. Frisch}
\affil{Dept. Astronomy and Astrophysics, University of Chicago,
Chicago, IL  60637}
\email{frisch@oddjob.uchicago.edu}

\author{Jacob Heerikhuisen}
\affil{CSPAR, University of Alabama, Huntsville, AL}
\email{jacobh@ucr.edu}

\author{Nikolai V Pogorelov}
\affil{CSPAR, University of Alabama, Huntsville, AL}
\email{nikolaip@ucr.edu}

\author{Bob DeMajistre}
\affil{Applied Physics Laboratory, Johns Hopkins University, Laurel, MD}
\email{Bob.DeMajistre@jhuapl.edu}

\author{Geoffrey B Crew}
\affil{Massachusetts Institute of Technology, Cambridge, MA}
\email{gbc@haystack.mit.edu}

\author{Herbert O Funsten}
\affil{Los Alamos National Laboratory, Los Alamos, NM}
\email{hfunsten@lanl.gov}

\author{Paul Janzen}
\affil{Dept. Physics Astronomy, University of Montana, Missoula, MT}
\email{paul.janzen@umontana.edu}

\author{David J McComas\altaffilmark{1}}
\affil{Southwest Research Institute, San Antonio, TX}
\altaffiltext{1}{University of Texas, San Antonio, TX}
\email{DMcComas@swri.edu}

\author{Eberhard Moebius}
\affil{Space Science Center, University of New Hampshire, Durham, NH}
\email{eberhard.moebius@unh.edu}

\author{Hans-Reinhard Mueller}
\affil{Dept. of Physics and Astronomy, Dartmouth College, Hanover, NH}
\email{Hans-Reinhard.Mueller@Dartmouth.edu}

\author{Daniel Brett Reisenfeld}
\affil{Dept. Physics Astronomy, University of Montana, Missoula, MT}
\email{dan.reisenfeld@umontana.edu}

\author{Nathan A Schwadron}
\affil{Boston University, Boston, MA}
\email{nathanas@bu.edu}

\author{Jonathan D. Slavin}
\affil{Harvard-Smithsonian Center for Astrophysics, Harvard, Cambridge, MA}
\email{jslavin@cfa.harvard.edu}

\author{Gary Paul Zank}
\affil{CSPAR, University of Alabama, Huntsville, AL}
\email{zank@email.cspar.uah.edu}

\begin{abstract}

The Interstellar Boundary Explorer (IBEX) spacecraft is providing the
first all-sky maps of the energetic neutral atoms (ENAs) produced by
charge-exchange between interstellar neutral \HI\ atoms and
heliospheric solar wind and pickup ions in the heliosphere boundary
regions. The 'edge' of the interstellar cloud presently surrounding
the heliosphere extends less than 0.1 pc in the upwind direction,
terminating at an unknown distance, indicating that the outer boundary
conditions of the heliosphere could change during the lifetime of the
IBEX satellite. Using reasonable values for future outer heliosphere
boundary conditions, ENA fluxes are predicted for one possible source
of ENAs coming from outside of the heliopause.  The ENA production
simulations use three-dimensional MHD plasma models of the heliosphere
that include a kinetic description of neutrals and a Lorentzian
distribution for ions.  Based on this ENA production model, it is then
shown that the sensitivities of the IBEX 1.1 keV skymaps are sufficient
to detect the variations in ENA fluxes that are expected to accompany 
the solar transition into the next upwind cloud.  Approximately 20\% of
the IBEX 1.1 keV pixels appear capable of detecting the predicted
model differences at the $ 3 \sigma$ level, with these pixels
concentrated in the Ribbon region.  Regardless of the detailed ENA
production model, the success of the modeled \BdotR\ directions in
reproducing the Ribbon locus, together with our results, indicate that
the Ribbon phenomenon traces the variations in the heliosphere
distortion caused by the relative pressures of the interstellar
magnetic and gaseous components.
\end{abstract}



\section{Introduction}

The dynamic heliosphere varies with the properties of the surrounding
interstellar cloud and the solar wind.  A theoretical study of the
expected heliosphere response to different types of interstellar
clouds show that both the overall dimensions and hydrogen filtration
should vary substantially with variations in the physical properties
of circumheliospheric interstellar material
\citep{Muelleretal:2006apj,MuellerWoodman:2008}. The presence of
an interstellar magnetic field causes heliosphere asymmetries
that can diagnose the properties of the surrounding interstellar material,
but which are partially offset by the charge-exchange coupling of interstellar
\HI\ and protons upstream of the heliopause \citep{Pogorelovetal:2009,OpherRichardson:2009,RatkiewiczJaffel:2008,Izmodenov:2009bfieldlism}.
The velocity discontinuity observed between interstellar gas inside of the
heliosphere \citep[e.g.][]{Witte:2004} and interstellar material (ISM)
towards the nearest stars in the upwind direction \citep[36 Oph and
$\alpha$ Cen,][]{Landsmanetal:1984,Lallementetal:1995,Wood36Oph:2000,LinskyWood:1996}
is usually interpreted to indicate that the Sun is immersed in one
interstellar cloud today, but will enter a separate cloud 
sometime in the next $\sim 4000$ years. The
Interstellar Boundary Explorer (IBEX) is for the first time mapping
heliospheric energetic neutral atoms, formed by charge-exchange
between solar wind ions and pickup ions with neutral interstellar
atoms \citep{McComasetal:2009ssr,McComas:2009sci,Funsten:2009sci,Fuselier:2009sci,Schwadron:2009sci}.  In this paper we
show that, depending on the source of ENAs observed, IBEX is capable
of detecting the variable heliosphere boundary conditions that might
accompany the expected (someday, it could be soon) solar transition
into a new interstellar environment in the upwind direction.

The discovery of an unexpected 'Ribbon' of ENA emission, in directions
where the interstellar magnetic field draping over the heliosphere is
thought to be perpendicular to the sightline, showed that IBEX may be
detecting plasma-neutral interactions beyond the heliopause.  The
similar spectra of ENAs in the Ribbon and adjacent sightlines suggest
that the Ribbon represents a selection effect rather than an ENA
population with an fully independent origin.  \citet{McComas:2009sci} and
\citet{Schwadron:2009sci} noted that the ribbon is organized by the
most likely direction of the external interstellar magnetic field
(ISMF), and proposed several different potential sources of the Ribbon
including the possibility that the Ribbon might be created from a
population of anisotropic suprathermal ions gyrating around the
interstellar magnetic field just outside the heliopause.  These ions
could be indigenous to the outer heliosheath (beyond the heliopause)
or more likely would arise from ENAs that propagated out from the
supersonic solar wind and/or inner heliosheath (between the
termination shock and the heliopause); these authors noted that the
problem with this idea is that the relatively confined pitch angle
distributions would need to be maintained long enough to create
"secondary ENAs", which likely takes several years on average.
Subsequently, \citet{Heerikhuisen:2010ribbon} incorporated this idea,
where outward propagating ENAs create an anisotropic population of
pickup ions (PUI) in the outer heliosheath that seed "secondary ENA" production
several hundred AU upwind of the heliopause, into a quantitative
model.  While it is still uncertain how long the ion ring beam
takes to scatter into a shell distribution (Florinski, Zank,
Heerikhuisen, Hu and Khazanov, submitted), with marginally stable ring
distributions predicted by some models for the distribution of the
pitch angles of pickup ions in the outer heliosheath (Gamayhunov,
Zhang, Rassoul, submitted to ApJ), the Heerikhuisen et al. simulation
assumes that the re-neutralization time is essentially instantaneous
compared to the scattering time so that the new secondary ENA will
have a preferred direction that is perpendicular to the local ISMF
direction.  In this model, IBEX then sees these secondary ENAs where
the gyration plane of the ion is aligned with the sightline to IBEX,
i.e.  where the sightline is perpendicular to the ISMF direction.

IBEX data are  uniquely
qualified to simultaneously test both the direction of the ISMF at the
heliosphere and the density of interstellar neutrals.
The ISMF drapes around the heliosphere,
 rotating by $\sim 30^\circ$ between 'infinity' and
the heliopause near the nose.   The
Ribbon is $\sim 20^\circ$ wide and at least $ 270^\circ$ long,
possibly forming a complete circle in the sky.  IBEX only sees ENAs
with momentum vectors directed back towards the inner heliosphere.
The long mean free paths of ENAs, e.g. $\sim 200$ AU for 1.1 keV ENAs
in \npro$\sim 0.1$ \cc\ plasma, allow detection in the inner
heliosphere of secondary ENAs formed in regions beyond the heliopause
with elevated interstellar densities and relatively isotropic ENA
velocities (compared to the outwards radial flows for primary ENAs
produced in the supersonic solar wind, although not compared to the
inner heliosheath ion populations).  The solar wind contributing
to ENA production includes both
core ions from the expanding solar corona, and pickup ions formed
inside of the heliosphere by charge exchange between interstellar
neutrals and the core solar wind. 

The predictive capabilities of global heliosphere models have improved
significantly to accommodate observational constraints placed by the
10 AU difference in the termination shock distances found by the
Voyager 1 and Voyager 2 spacecraft \citep{Stone:2007,OpherRichardson:2009}, the $\sim
5^\circ$ offset between the upwind directions of interstellar \HI\ and
\HeI\ flowing into the heliosphere determined from SOHO/SWAN and
Ulysses/GAS data \citep[][where the \HeI\ upwind direction must first
be converted to J2000 coordinates for this
comparison]{Lallementetal:2005,Witte:2004,Frisch:2008s1}, the
properties of the ISM surrounding the heliosphere
\citep{SlavinFrisch:2008}, and now the IBEX data on ENA fluxes and the
Ribbon.  Although the IBEX Ribbon itself was not predicted by models
of ENA formation in the heliosphere, the global heliosphere models
provided the ISMF orientation that matches well with the configuration
of the Ribbon in the sky \citep{Schwadron:2009sci,Pogorelovetal:2009}.

In the discussions below we rely on the ENA production models
quantified by
\citet{Heerikhuisen:2010ribbon,Pogorelovetal:2009strongismf} to
predict the ENA fluxes for a heliosphere immersed in the next
interstellar cloud versus the present-day surrounding interstellar
cloud.  The ENA production simulations use
three-dimensional MHD plasma models of the heliosphere that include a
kinetic description of neutrals and a Lorentzian distribution for
solar wind protons to approximate the suprathermal population of
pick-up ions in the heliosheath region.  The interstellar neutrals,
that are coupled self-consistently to the plasma component in the MHD
heliosphere models, act to symmetrize the heliosphere.  Any asymmetry
in the quiescent plasma distribution (e.g. created by ISMF) results in
variations in the number of charge-exchange events, creating new ions
that are decelerated by the heliopause and therefore that mitigate the
asymmetry \citep{Pogorelovetal:2008let,Pogorelovetal:2009,
Pogorelovetal:2009strongismf}.  The LISM flow, for the assumed ISMF strength, is
subfast magnetosonic (Table 1).  This results in the absence of a bow
shock in front of the heliopause, and increases the width of the
region where the ISMF deviates from its unperturbed orientation.
Since the local ISMF direction varies as interstellar protons approach
the heliopause, the mean-free-path for the charge exchange interaction
must also be included self-consistently in any Ribbon production
models. 

The Ribbon
ENAs are formed upwind of the heliopause in the
\citet{Heerikhuisen:2010ribbon} model, so the outer boundary
conditions set by the ISM have a direct impact on predicted ENA fluxes
and provide a means of estimating ENA variations from a cloud
transition, regardless of whether the model is correct in detail.  Any
model that reproduces the observed \BdotR\ of the Ribbon,
which is seen where the ISMF ($ {\bf B}$) is perpendicular to the sightline
(${\bf R}$), should provide useful insights into the deformation of the
heliosphere due to altered boundary conditions from the variable
interstellar wind.  Our conclusions here rely explicitly on the
assumption that the \citet{Heerikhuisen:2010ribbon} model provides a
viable description of ENA production, both for the cloud we are in
today and for the nearby cloud observed towards $\alpha$ Cen and 36
Oph in the upwind direction.

We focus on the 1.1 keV data, because the contrast between Ribbon ENA
fluxes and diffuse ENA fluxes is stronger at this energy, and partly
because of the enhanced outwards flow of ENAs from core solar wind
ions that have typical energies near 1 keV.  The ENA spectra are
explicitly predicted by the heliosphere models, however this spectral
information is not used here.  The observed fluxes of $\sim 4.5$ keV
ENAs are an order of magnitude lower than the 1.1 keV fluxes and the
mean-free-paths are $\sim 50$\% larger. Since the outflowing ion
fluxes will decrease as $\sim R^{-2}$ with distance from the Sun $R$,
both parent ion densities and the resulting 1-AU ENA densities are
lower when production regions are further from the Sun.

\section{Properties of the Upwind ISM  \label{sec:ism}}

The boundary conditions of the heliosphere are set by the ISM, and
 vary over geologically short time-scales.  Interstellar clouds within
 $\sim 50$ pc flow past the Sun with heliocentric velocities that
 cluster around $\sim 28$ \kms\ (after projection effects are
 removed), or $\sim 1$ parsec per 35,000 yrs
 \citep{FrischSlavin:2006astra}.  If nearby ISM is in pressure
 equilibrium, then models of the interstellar cloud around the
 heliosphere \citep{SlavinFrisch:2008} combined with ISM data
 \citep{RLIII} yield an estimate for the typical cloud length of $\sim
 1$ pc.  Cloud column densities for 23 cloud components towards stars within
 10 pc yield a range of cloud lengths 0.06--3 pc, giving typical cloud
 crossing times for the Sun of $\sim 1,450 - 2.8 \times 10^5$
 years.\footnote{These estimates assume an equilibrium thermal
 pressure of $\sim 3 \times 10^{-13} $ dynes \cmtwo\ for the
 present-day cloud, uniform magnetic and cosmic ray pressures,
 D/H$\sim 1.6 \times 10^{-5}$, and a uniform proton density of 0.08
 \cc}.  

The velocity of the ISM at the heliosphere, which we term here
 the heliospheric ISM but which is also known as either the Local Interstellar
Cloud (LIC) or the
 circumheliospheric ISM, is best set by the velocity of interstellar
 \HeI\ observed inside of the heliosphere by the Ulysses satellite
 \citep{Witte:2004}.  Interstellar \HeI\ experiences minimal
 filtration in the heliosheath regions
 \citep[$<2$\%,][]{MuellerZankfilt:2004} and is not subject to
 radiation pressure, so that variation in the interstellar \HeI\
 velocity as it traverses the heliosphere is minimal.  The first
 comparisons between the velocities of interstellar gas in the
 heliosphere and outside of the heliosphere, towards stars in the
 upwind direction, showed differences of over 3 \kms\
 \citep{AdamsFrisch:1977}.  When the 26.3 \kms\ \HeI\ velocity is
 projected towards the nearest star in the upwind direction $\alpha$
 CMa, 1.3 pc away and 50$^\circ$ from the heliosphere nose, the
 projected heliospheric ISM velocity is --17.0 \kms, in contrast to
 the observed cloud velocity from the unsaturated \FeII\ and \MgII\
 lines of $- 18.0 \pm 0.2$ \kms\ \citep{LinskyWood:1996}. The
 heliosphere nose direction is given by the upwind direction of \HeI\
 flowing through the heliosphere, or $\lambda,\beta=255.4^\circ \pm
 0.5^\circ,+5.1 \pm 0.2^\circ $ \citep[epoch J2000,][and private
 communication]{Witte:2004}.  The star 36 Oph is 6 pc away and
 10$^\circ$ from the heliospheric nose.  The projected heliospheric
 ISM velocity in this direction is --25.9 \kms, versus the observed
 cloud velocity from the \FeII\ and \MgII\ lines of $-28.1 \pm 0.2$
 \kms\ \citep{Wood36Oph:2000}.  The limit on a cloud component at the
 heliospheric ISM velocity towards 36 Oph is $\sim 6 \times 10^{16}$
 \cmtwo, giving an upper limit to the heliospheric ISM
 edge\footnote{We use \nHI=0.19 \cmtwo\ for the heliospheric ISM \HI\
 density, based on Model 26 in the radiative transfer models of
 \citet{SlavinFrisch:2008}.} in this direction of 0.1 pc which will be
 traversed in less than 4000 years.  

The interstellar cloud observed
towards 36 Oph and $\alpha$ Oph is known as the G-cloud
 \citep{LallementBertin:1992,FGW:2002,RLIV:2008vel}, since it is close
 to the Sun in the galactic hemisphere.  Although other possible
 clouds have been suggested as the next cloud to be encountered by the
 heliosphere \citep{Frisch:2003apex}, the G-cloud remains the most
 likely future heliospheric environment.

The scenario examined here assumes that the Sun transitions directly
from the heliospheric ISM to the G-cloud seen towards 36 Oph and
$\alpha$ Cen.  We determine the properties of the next cloud by
assuming that the thermal and magnetic pressures in the heliospheric
ISM are equal, and that the heliospheric ISM and G-clouds are in
pressure equilibrium.  The G-cloud temperature is found from the
mass-dependent Doppler broadened widths of interstellar absorption
lines, and is $5400 \pm 500$ K towards $\alpha$ Cen, and $5900 \pm
500$ towards 36 Oph (where the cloud column density is also larger by
70\%).  The cooler G-cloud temperatures are thus compensated by
neutral densities that are 20\% larger than the heliospheric ISM.  The
9\% higher velocity of the G-cloud, in comparison to the velocity of
the ISM now surrounding the Sun, will increase
the interstellar ram pressure even if thermal pressure and ionization levels
remain constant.  We test the sensitivity of ENA emission to the ISMF
direction using two separate ISMF directions for the G-cloud.  The
first assumption is that the directions of the ISMF in the
heliospheric ISM and G-clouds are the same (Models 2 and 3 in Table
1).  For Model 4 we make an arbitrary\footnote{Caveat: This direction,
which is arbitrary from an interstellar viewpoint, was selected to lie
on the great circle that divides the hot and cold poles of the cosmic
microwave background dipole moment, and which passes through the
heliosphere nose region \citep{Frisch:2010s1}.}  assumption for the
G-cloud ISMF direction, which is $28^\circ$ different from today's
field but less than the $\sim 30^\circ$ rotation of the ISMF between the
ISM and the heliopause for the upwind direction.  
The detailed heliosphere boundary conditions used here for the next-cloud are
listed in Table 1.  
We show below that the IBEX Ribbon is highly
sensitive to even small variations in the ISMF direction, even when
increased \HI\ densities mitigate the influence of the ISMF on
heliospheric asymmetries.

\section{ENA fluxes from encounter with Next Interstellar Cloud}

\begin{figure}[hb!]
\plotone{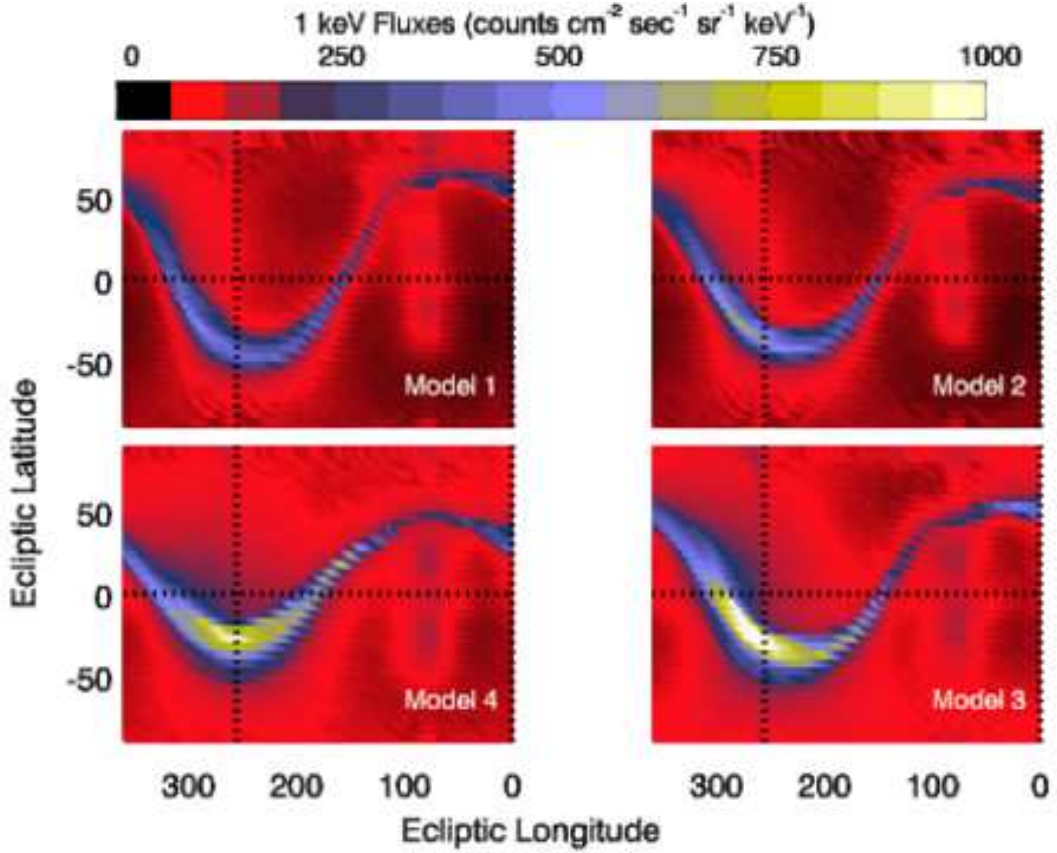}
\caption{ Four models of ENA fluxes at 1.1 keV, based on different
values for the ISM that constrains the heliosphere (see Table 1).
Model 1 is the initial model used to evaluate the IBEX results
\citep{Schwadron:2009sci}.  Model 2 is the assumed benchmark model for
the production of ENAs observed by IBEX today (Heerikhuisen et
al. 2010).  Models 3 and 4 utilize the same model, but with different
heliosphere boundary conditions appropriate for the next upwind cloud.
Model 3 represents 1.1 keV ENAs for a heliosphere constrained by the
physical parameters of the next cloud in the upwind direction, where
it is assumed that cloud is in pressure equilibrium with the
circumheliospheric gas.  The same ISMF direction of
$\lambda,\beta=224^\circ,41^\circ$ is assumed for Models 2 and 3.
Model 4 is the same as Model 3, except that the direction of the ISMF
differs by $28^\circ$, and is directed towards
$\lambda,\beta=252^\circ,42^\circ$.  The dashed lines, in this and
subsequent figures, intersect at the longitude of the heliosphere nose
($\lambda=255.4^\circ $) and the ecliptic plane.}
\label{fig:4models}
\end{figure}

The heliosphere model has been run for the interstellar boundary
conditions listed for Models 1-4 in Table 1.  The 1.1 keV ENA fluxes
predicted by the resulting models are displayed in
Fig. \ref{fig:4models}.  Model 1 corresponds to the heliosphere model
displayed in \citet{Schwadron:2009sci}.  Model 2
\citep[from][]{Heerikhuisen:2010ribbon} is the updated model we use
for the heliosphere today. Models 3 and 4 correspond to the
anticipated heliosphere environment in the next interstellar cloud,
with Model 3 having a similar ISMF configuration as Model 2, and Model
4 showing a ISMF field with a different direction.  The increased \HI\
density in Model 3 produces a brighter ribbon, while the ram and
thermal pressures slightly increase and the magnetic pressure slightly
decreases.  The magnetic field in Model 2 plays a bigger role in
deforming the heliosphere than in Model 3.  Model 4 has the Ribbon
shifted significantly because of the different ISMF direction.

\begin{figure}[hb!]
\begin{center}
\plottwo{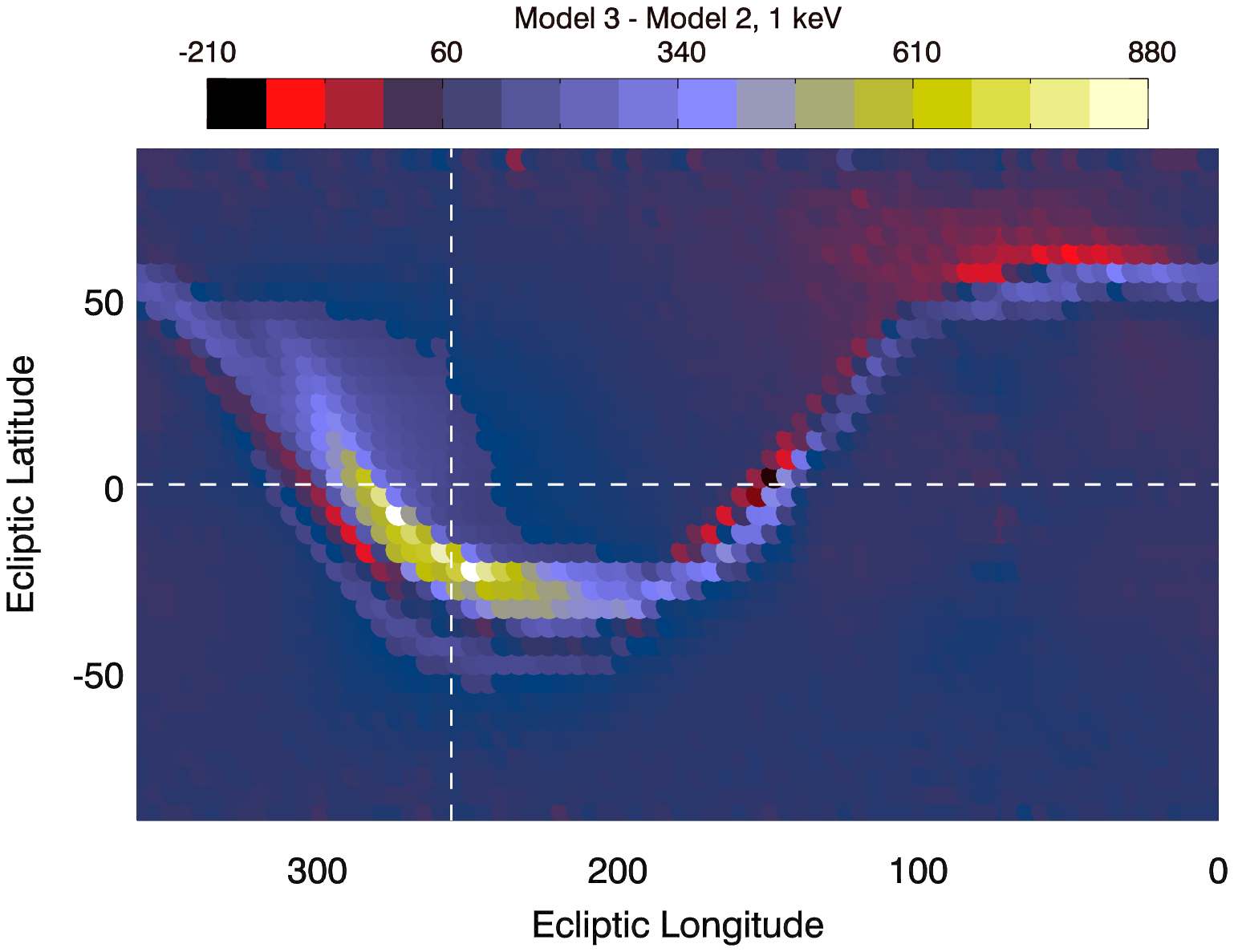}{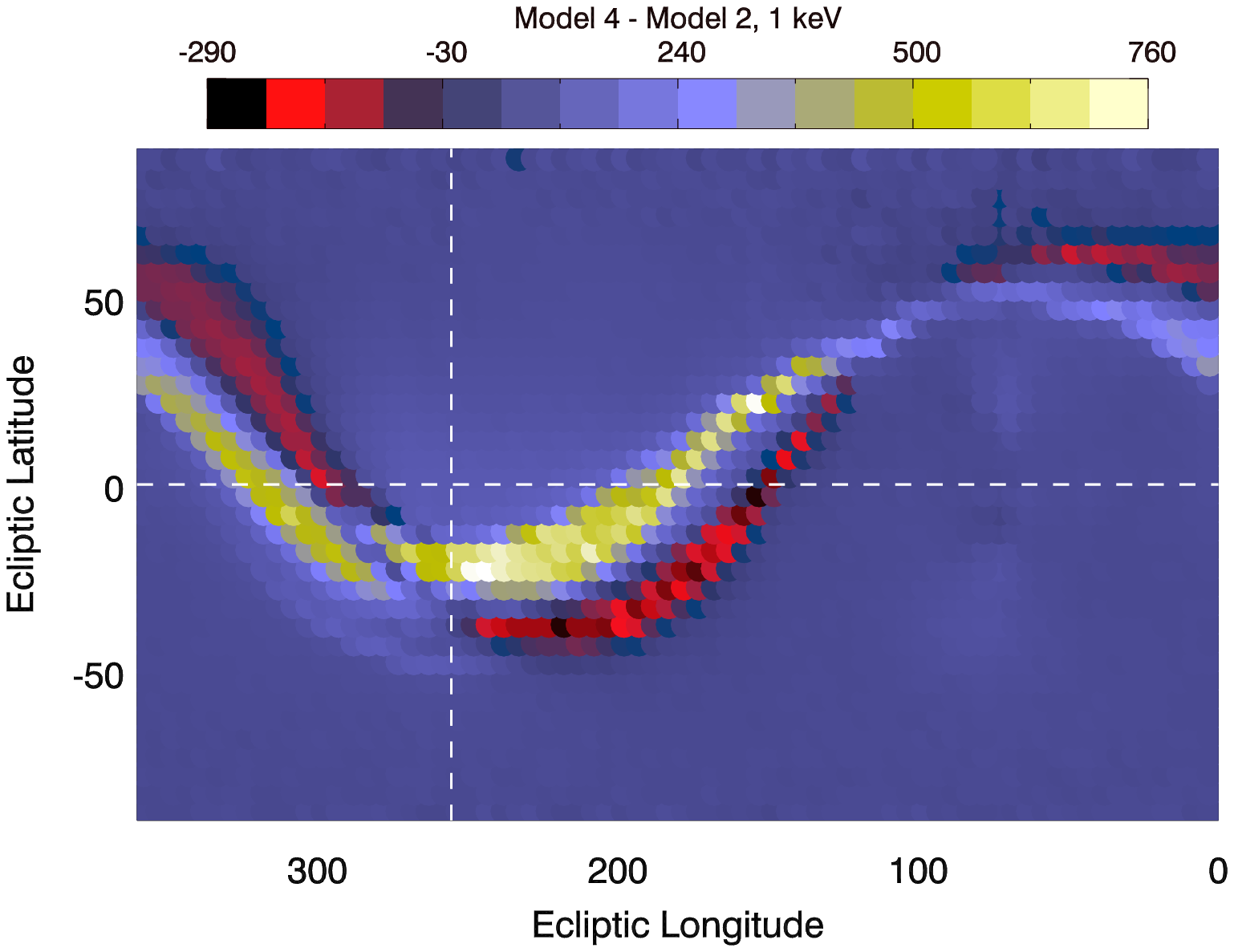}
\end{center}
\caption{The differences between the ENA fluxes produced by the
nominal G-cloud and the nominal LIC.  Left: Model 3--Model 2
(\Fmodthree -- \Fmodtwo).  Right: Model 4--Model 2 (\Fmodfour --
\Fmodtwo).  The different distortions of the heliosphere caused by
variations in the relative magnetic and thermal ram pressures of the
ISM are apparent on the northeast flank of the heliopause in the left
figure.  The right figure shows that the ribbon configuration is
highly sensitive to variations in the ISMF direction. The differences
in the distortion of the heliosphere towards the tail region are
marginally visible.  Fluxes are given in the units of \counts.  }
\label{fig:moddiff}
\end{figure}

The differences between the predicted ENA fluxes for Model 2 (the
'today-cloud') and Model 3 (the G-cloud assuming the same ISMF
direction as today-cloud) are directly displayed in
Fig. \ref{fig:moddiff}.  The ENA flux differences are obtained by
subtracting the predicted fluxes of Model 2 (\Fmodtwo) from the
predicted fluxes of Model 3 (\Fmodthree, left).  The most significant
difference between the two models is the ram pressure of the neutrals,
which is a factor of 1.8 larger for Model 3 versus Model 2.  Over most
of the sky, the higher flux of interstellar \HI\ into the heliosphere
for Model 3 generates larger ENA fluxes, with the differences
approaching the brightest observed fluxes.  However the red pixels
in Fig. 2, left, show regions where the today-cloud has higher fluxes
than the next-cloud, and represent the small shift in the Ribbon
position due to the increase in the ratio of thermal to magnetic
pressures, \Pthr/\Pmag, in Model 3 compared to Model 2.  The effect of
increased \HI\ densities and thermal pressures are also seen in the
increased ENA emissivity of the eastern flank of the heliosphere in
the nose direction, where there is a bulge in total pressure
\citep[magnetic and thermal, see Fig. 1 in
][]{Heerikhuisen:2010ribbon}.  Fig. \ref{fig:moddiff}, right, shows
the flux differences between Model 2 and Model 4, where the ISMF
direction in the G-cloud has been allowed to vary by 20$^\circ$. This
modest variation in the ISMF direction, while retaining the same field
strength, leads to an obvious shift in the Ribbon location, and shows
that variations in the direction of the ISMF draping over the
heliosphere should be apparent in the ENA data.

\begin{figure}[hb!]
\begin{center}
\plottwo{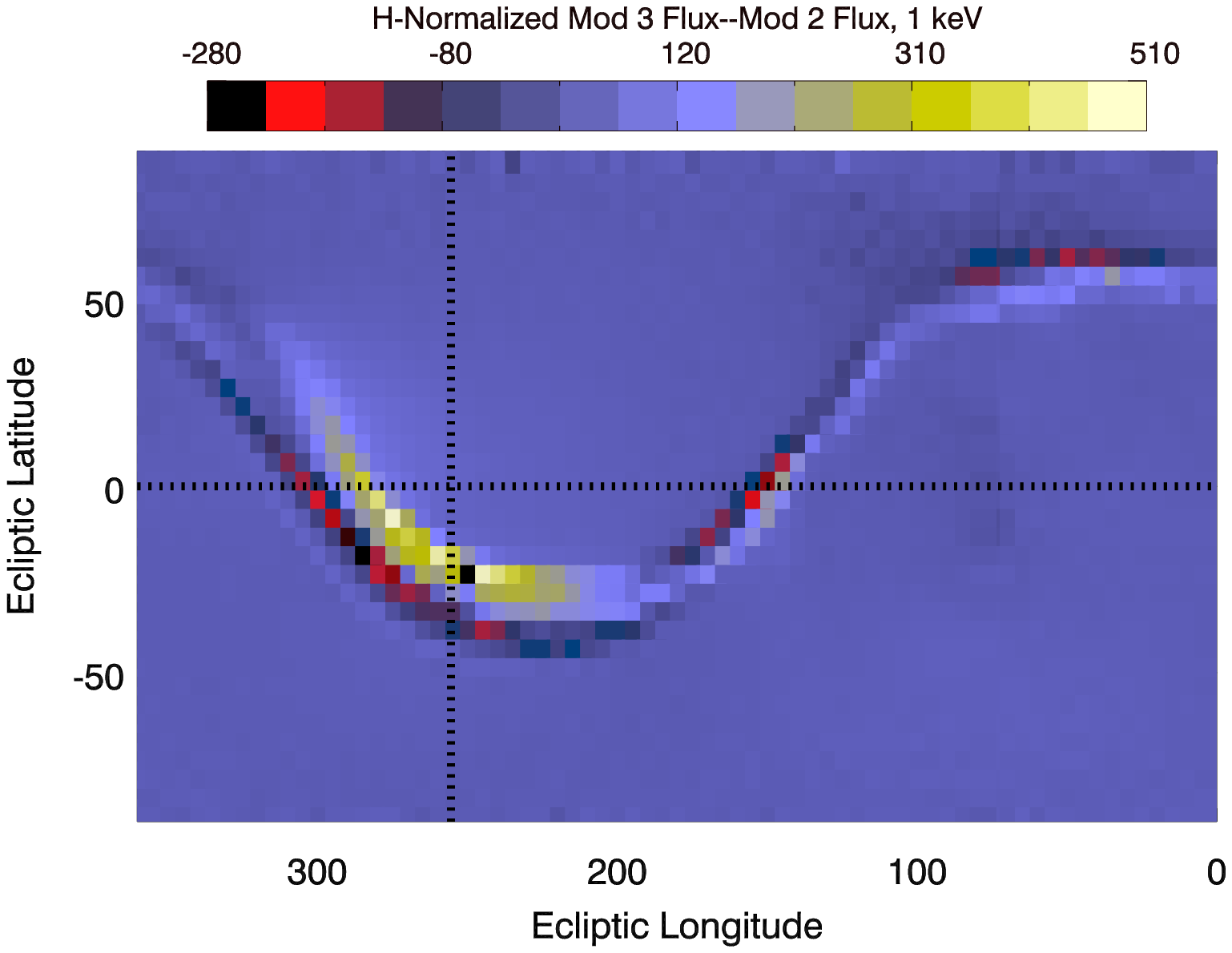}{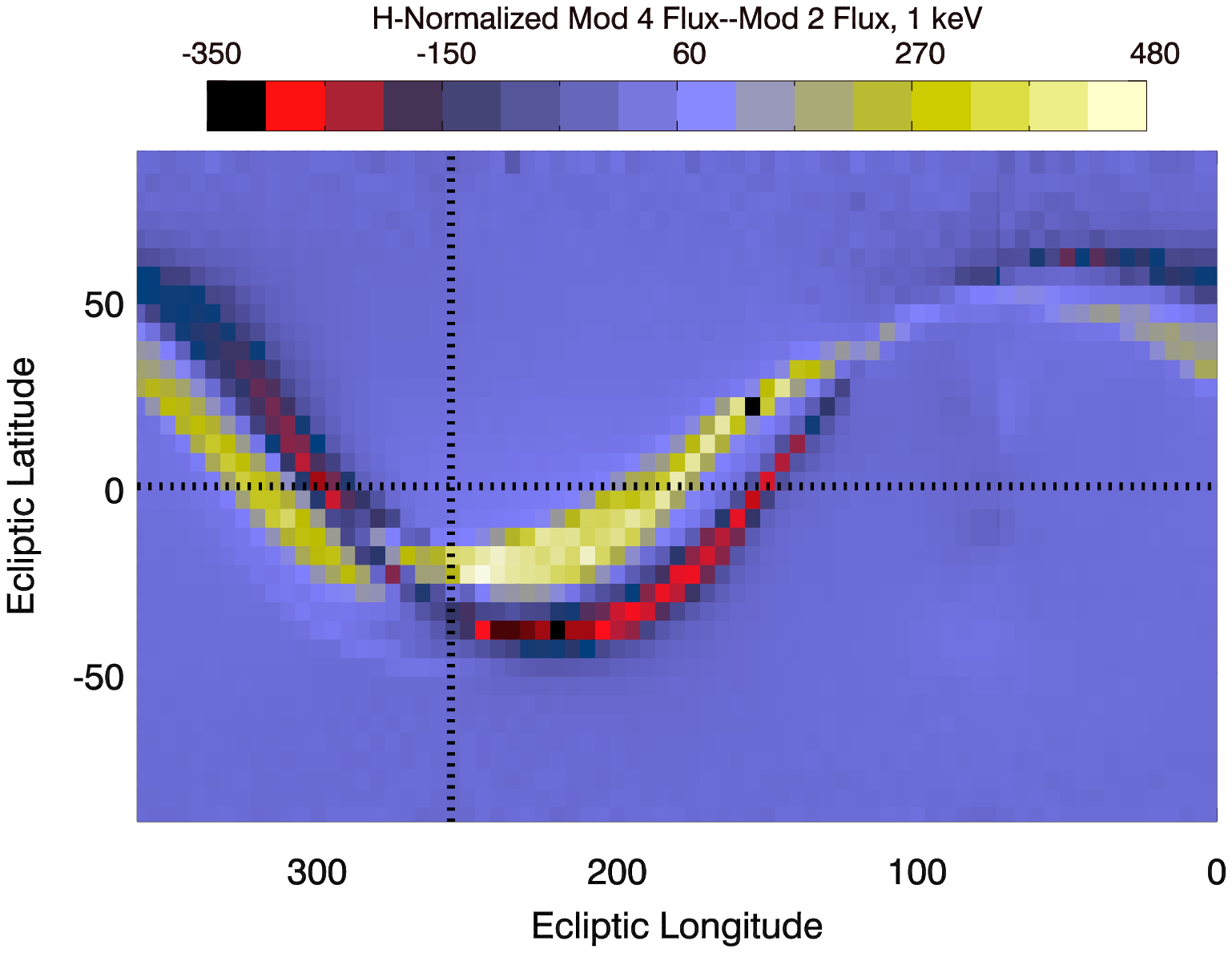}
\end{center}
\caption{The differences between the ENA fluxes produced by the
nominal G-cloud and the nominal present-day cloud, after normalizing
the fluxes in Models 3 and 4 by the ratio of the neutral densities (0.65).
Left: (\nHImodtwo/\nHImodthree)*\Fmodthree -- \Fmodtwo.  Right:
(\nHImodtwo/\nHImodfour)*\Fmodfour -- \Fmodtwo.  Zero values of the
difference (blue) along the ribbon imply that the ribbons in the two
models overlay each other. The background blue regions, where
difference counts are $\sim 0$, indicate the regions where ENA fluxes are
linearly related to the neutral interstellar density in these
models.  Fluxes are given in the units of \counts.}
\label{fig:moddiff2}
\end{figure}

The ENA-production model used here predicts that the ribbon 
and non-ribbon regions respond differently to variations in the
interstellar density, because the ribbon also directly traces 
heliosphere asymmetries created by the ISMF.
Fig. \ref{fig:moddiff2} shows the model differences modified by the ratio
of \nHI\ in Models 2 and 3 (0.65).
The left figure shows 0.65 *\Fmodthree -- \Fmodtwo, and the
right figure shows 0.65*\Fmodfour -- \Fmodtwo.  
The background blue regions, where difference counts are
$\sim 0$, indicate regions where the ENA fluxes are linearly related to the
neutral interstellar density.  The variations in the ribbon colors show that
the ribbon ENAs trace the distortion of the heliosphere, which depends
on the asymmetries introduced by the relative interstellar, ram, and
thermal pressures, and which changes with different ISM conditions.

\begin{figure}[hb!]
\begin{center}
\plottwo{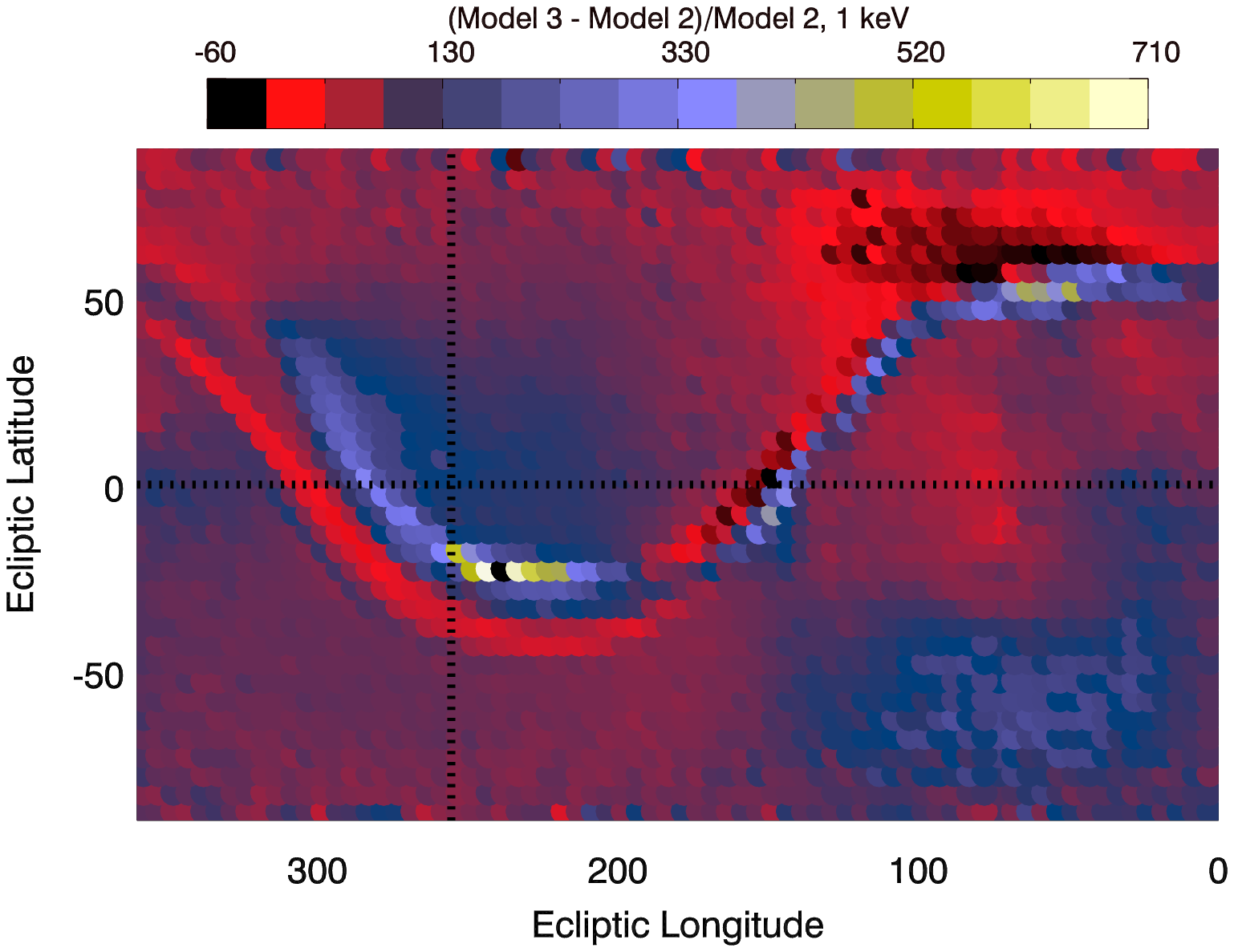}{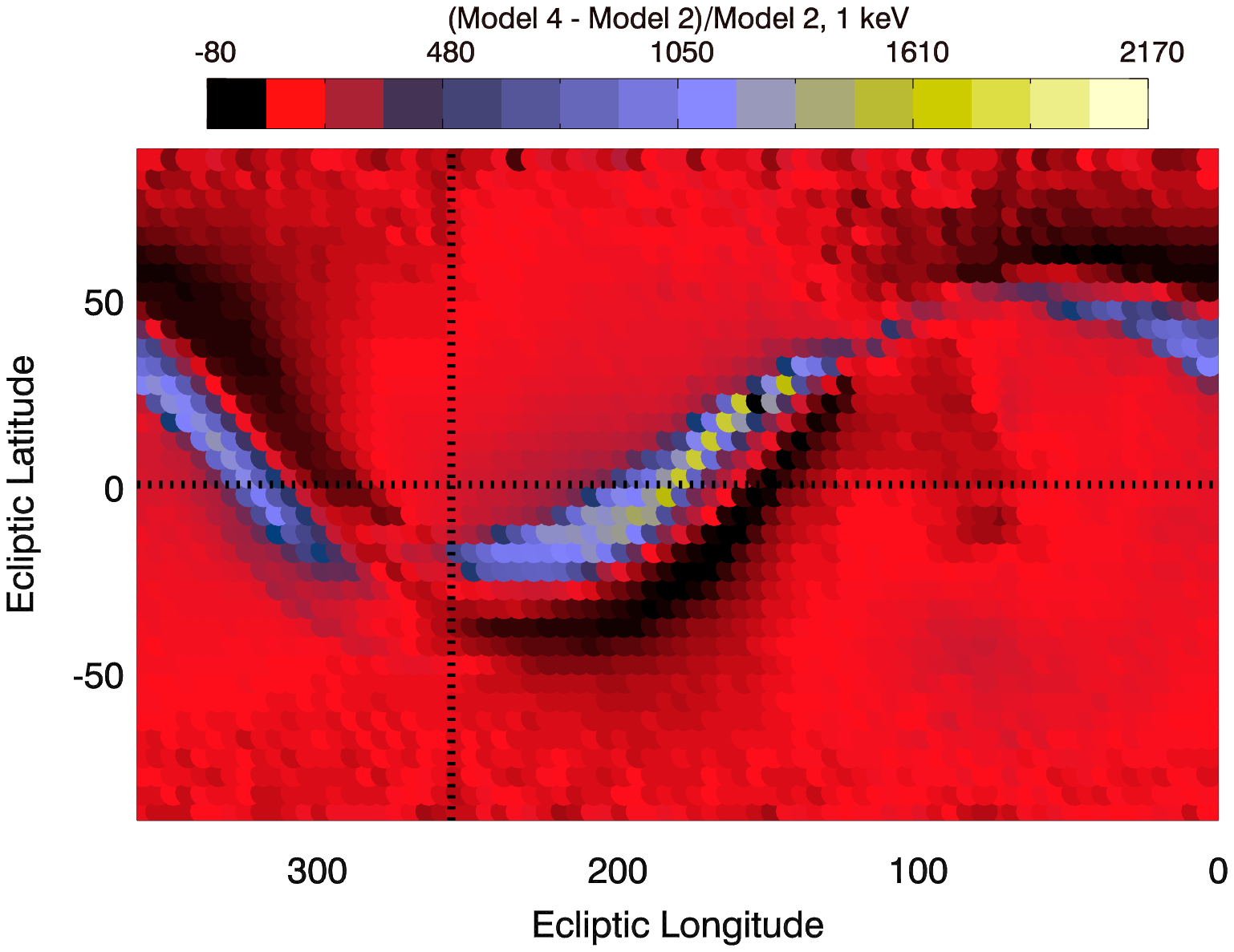}
\end{center}
\caption{The percentage differences between the 1.1 keV ENA fluxes
produced by the nominal G-cloud and the nominal present-day cloud.
Left: 100(\Fmodthree--\Fmodtwo)/\Fmodtwo.  Right: 100(\Fmodfour--\Fmodtwo)/\Fmodtwo.
The right figure shows that the ribbon configuration is highly
sensitive to any variation in the ISMF direction.  }
\label{fig:percent}
\end{figure}

The differences in the ENA fluxes predicted by the two next-cloud
models are quite obvious in the high-flux regions of the Ribbon, but
less obvious (for this color scale) for the directions towards the
tail.  In order to emphasize the differences in the weaker diffuse ENA
emission originating in the low-flux tail region, Fig. \ref{fig:percent}
shows the percentage
differences between Models 3 and 2, (\Fmodthree--\Fmodtwo)/\Fmodtwo,
and Models 4 and 2, (\Fmodfour--\Fmodtwo)/\Fmodtwo, with enhanced
color scales.  The percentage differences in the tail region
for Model 3 are larger than for Model 4. The outer heliosheath region
around the tail is the most disturbed part of the model results, since
the flows are subsonic.  Hence larger relative variations in ENA
fluxes are possible.

\begin{figure}[hb!]
\begin{center}
\plottwo{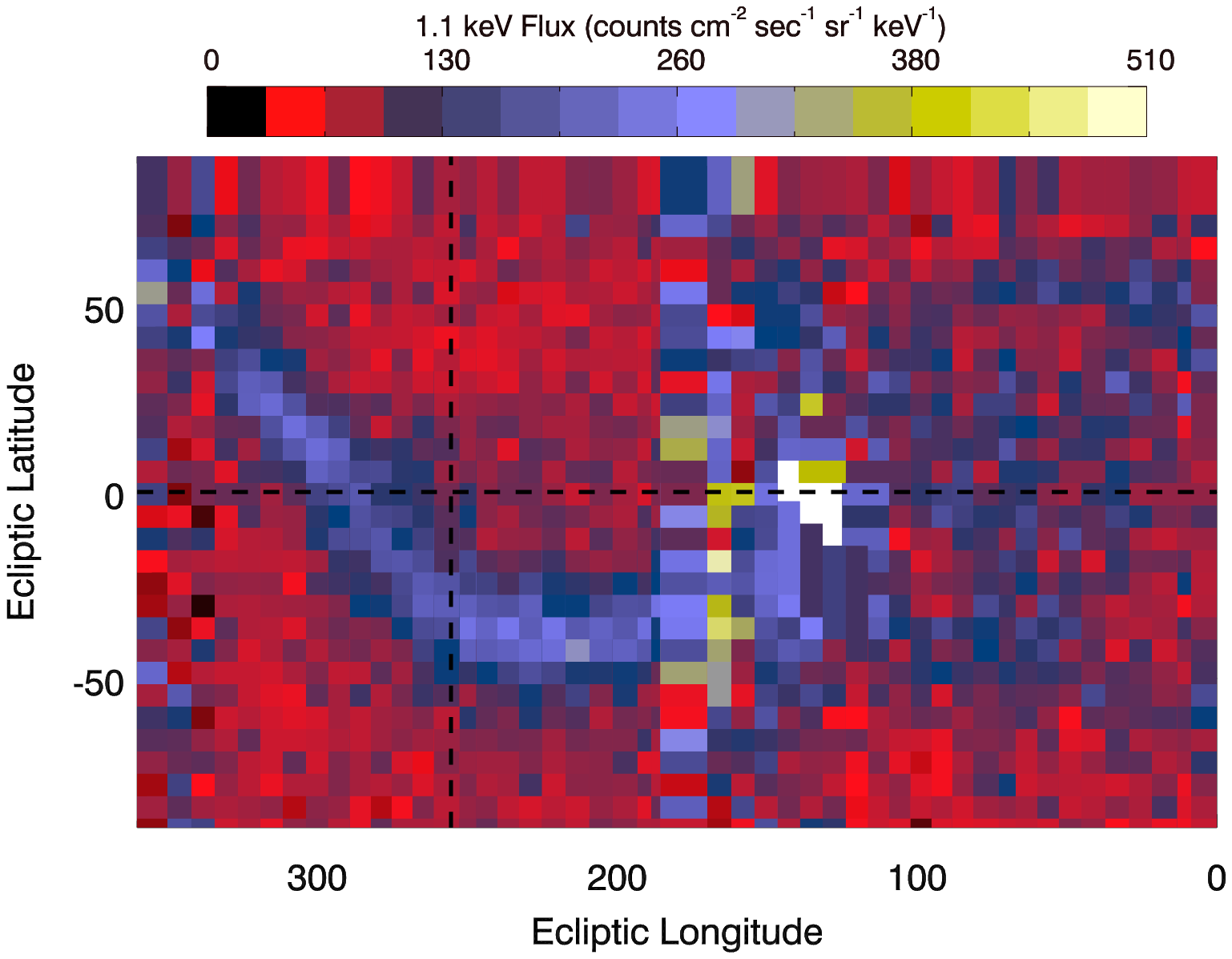}{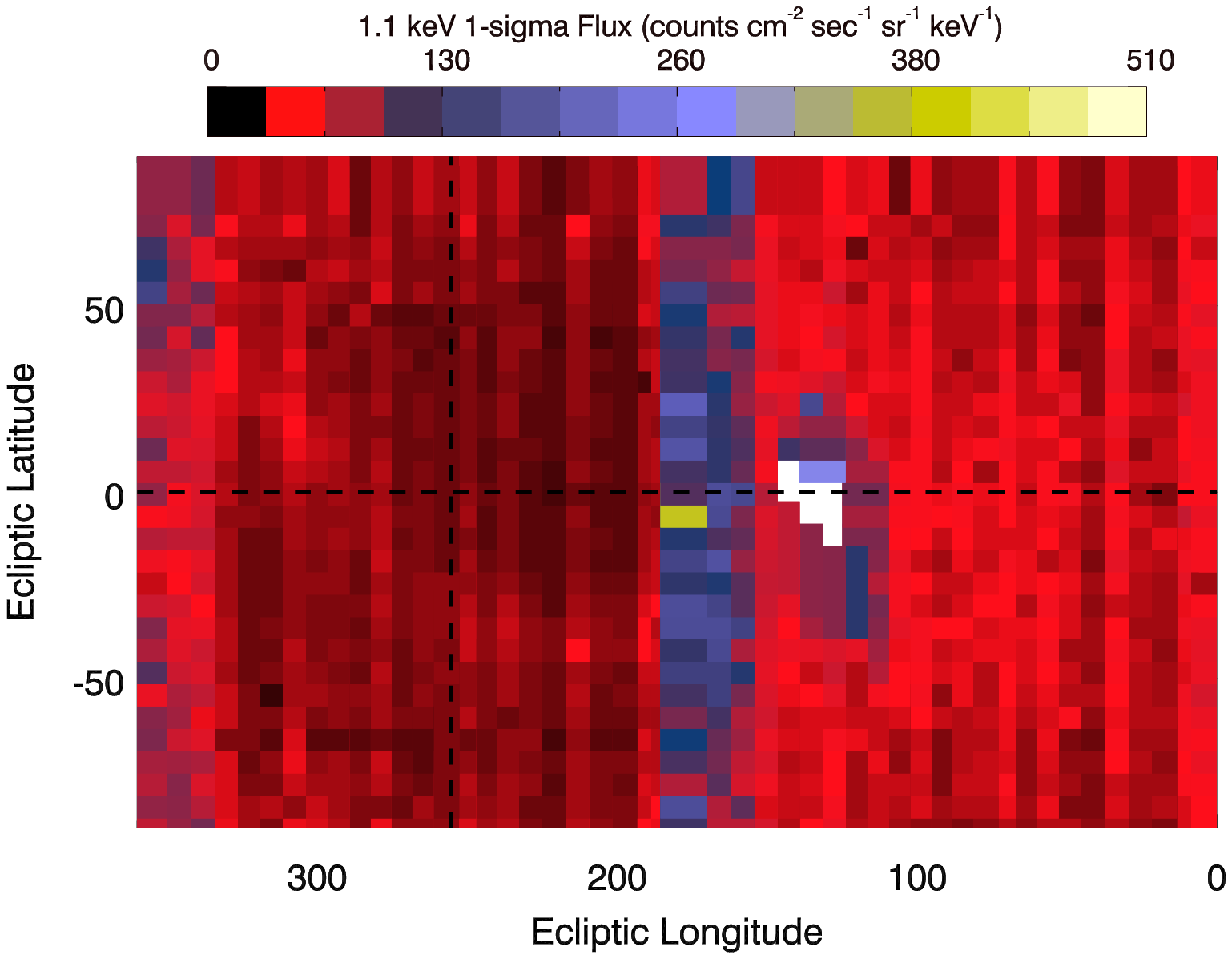}
\end{center}
\caption{Left: The first 1.1 keV IBEX ENA skymap, corrected for the
Compton-Getting effect (see text).  Right: The $1\sigma$ statistical
uncertainties on these data (see text).}
\label{fig:esa3}
\end{figure}

The capability of IBEX to detect the ENA variations shown in
Figs. \ref{fig:moddiff}--\ref{fig:percent} rests on the 
predicted differences in the modeled ENA fluxes 
compared to the flux uncertainties for the IBEX data.  For this
comparison, we use the first 1.1 keV ENA flux maps in the third
energy passband ("ESA 3") of the six energy passbands in the IBEX-HI
neutral atom imager \citep{Funstenetal:2009}, where fluxes are an
order of magnitude larger than at 4.5 keV (ESA 6).
Fig. \ref{fig:esa3}, left, shows ESA3 fluxes, after correction for the
Compton-Getting (CG) shift in the energy and spatial distribution of
high velocity particles due to the 30 \kms\ orbital motion of the
Earth \citep[e.g.][]{GleesonAxford:1968}.\footnote{We have used the
IBEX Compton-Getting corrected data set
"flxset\_hd60-id-base-0071-2010-04-09.sav", that is available at the
IBEX Science Operations Center (ISOC).} IBEX pixels are $\sim 7^\circ$.
The CG corrections are based on a
power-law energy spectra that are derived from adjacent energy
passbands, typically $\sim E^{-1.6}$, which is evaluated over the
look-direction and convolved with the energy response of the detector
(see the Appendix in McComas et al., 2010, submitted to GRL, for
details on the CG correction to the ENA energies measured by IBEX).
The $1\sigma$ uncertainties (\dFesa3) on these fluxes have been determined from
the Poisson statistics propagated from the measurement uncertainties.
The IBEX data are built from data processed by the IBEX Science
Operations Center (ISOC) for the first all-sky IBEX map (orbits
11-34).  The uncertainties in the ESA3 fluxes are shown in
Fig. \ref{fig:esa3}, right.

\begin{figure}[hb!]
\plottwo{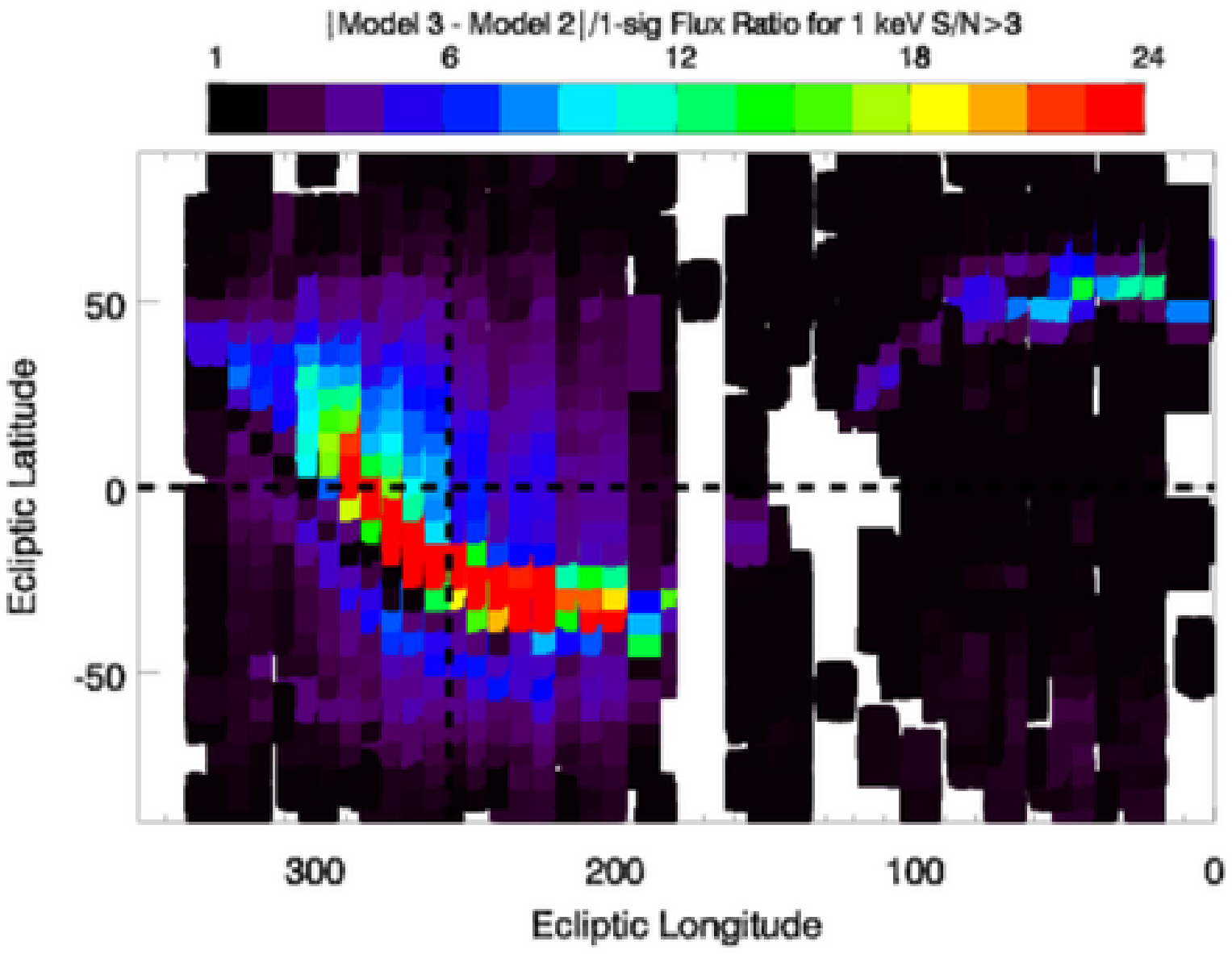}{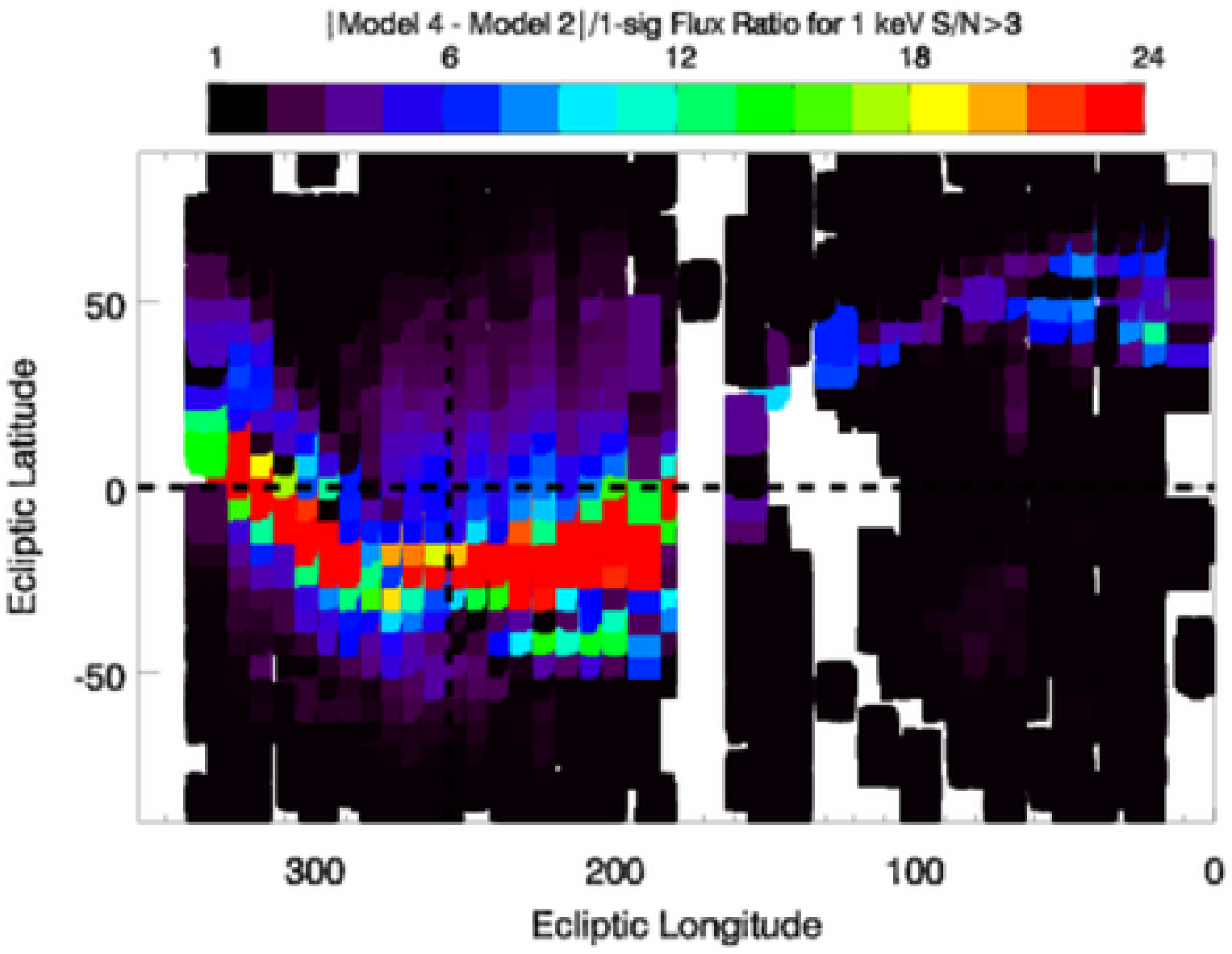}
\caption{The differences between ENA fluxes predicted by models
constrained by different boundary conditions, and compared to the $1
\sigma$ measurement uncertainties in the ESA 3 energy passband (1.1
keV, see text).  Left: The ratio of the absolute value of Model 3 minus Model 2
fluxes, to $1 \sigma$ uncertainties (\abs\Fmodthree
--\Fmodtwo\abs)/\dFesa3).  Right: Same as left figure, but showing the
absolute value of Model 4 minus Model 2 fluxes (\abs\Fmodfour
--\Fmodtwo\abs)/\dFesa3).  Pixels are left blank where fluxes are
insignificant (S/N$<3$).}
\label{fig:moddatdiff}
\end{figure}

\begin{figure}[hb!]
\plottwo{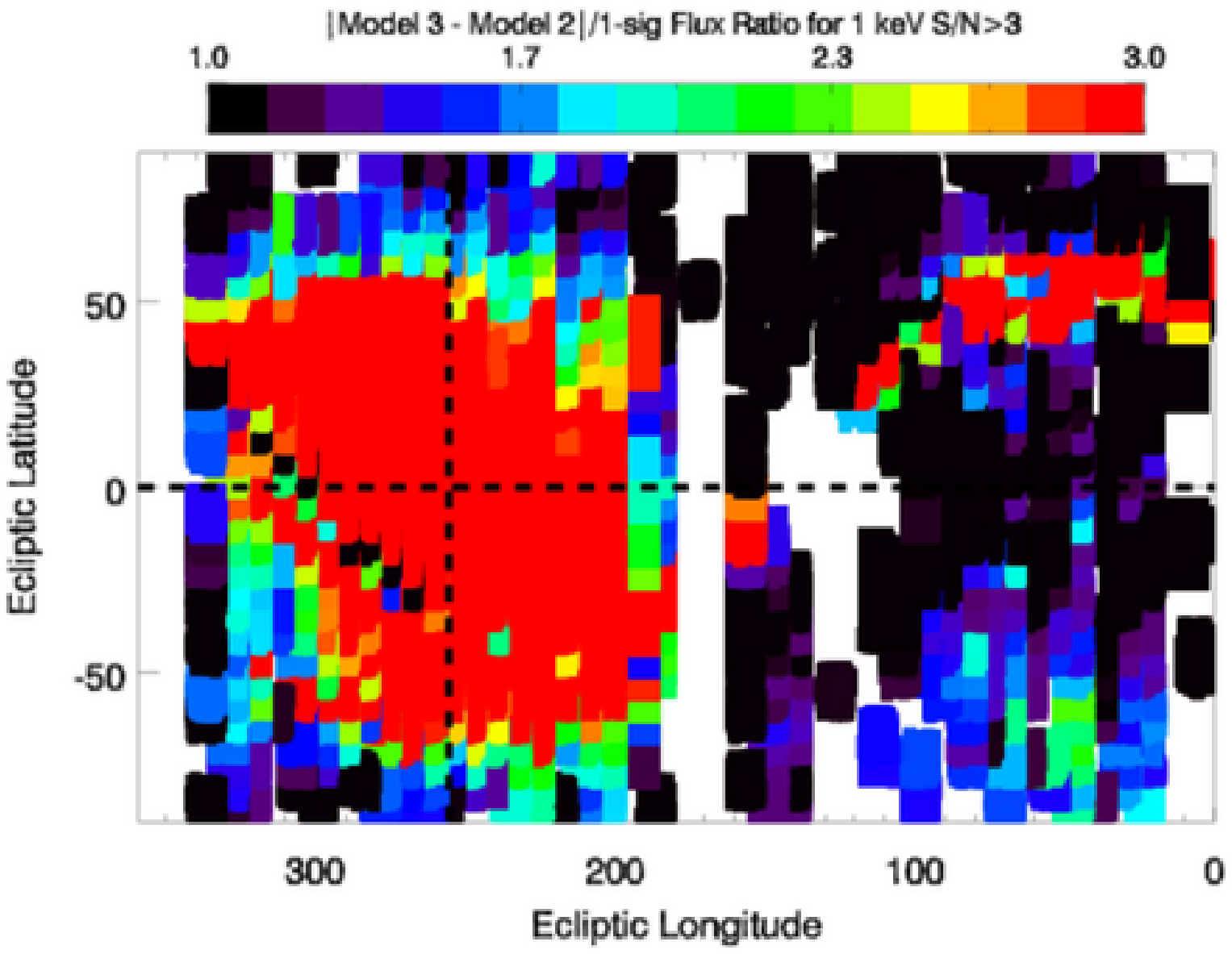}{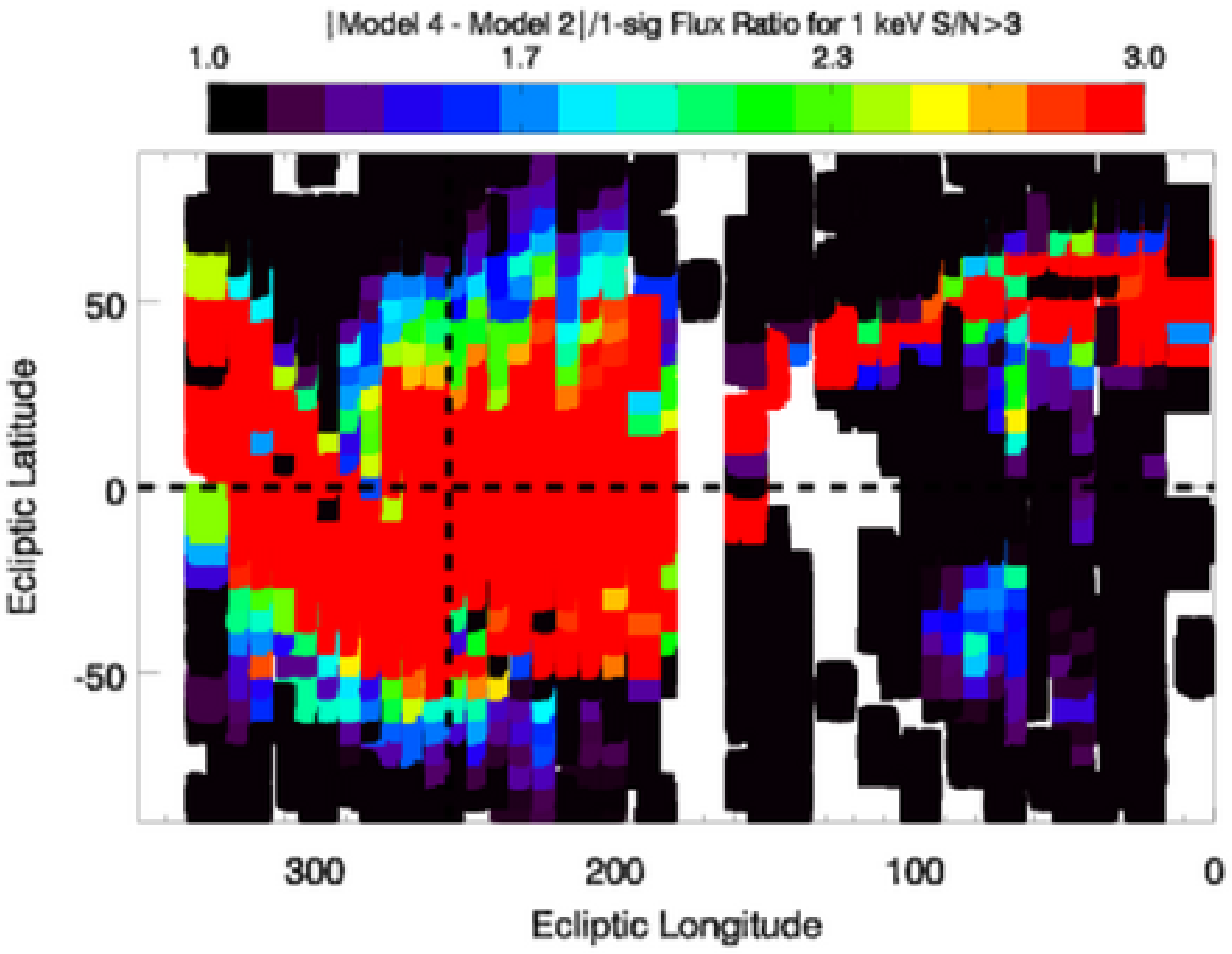}
\caption{Same plot as Fig. \ref{fig:moddatdiff}, except that the color
scale is coded to enhance the low flux areas in the tail region.}
\label{fig:moddatdifftail}
\end{figure}

The measurement uncertainties for 1.1 keV fluxes can be compared to
the  predicted ENA flux differences for the next-cloud versus
the present cloud, e.g.  \delmod=\abs\Fmodthree--\Fmodtwo \abs,
based on the  \citet{Heerikhuisen:2010ribbon} model.  For
the comparison we preselect data points with signal-to-noise S/N$>
3$.  In Fig. \ref{fig:moddatdiff}, individual pixels in the
Ribbon region for both models have values of \delmod/\dFesa3$ >> 5 $.
The same difference map is plotted in Fig. \ref{fig:moddatdifftail},
but now color-coded to enhance the differences in the tail region.
Lower ENA fluxes towards the tail yield \delmod/\dFesa3$ \sim 1-2 $
for individual pixels, which is somewhat larger for Model 3 than for
Model 4.  Groups of 25 pixels would yield a factor of 5 improvement in
the S/N of the difference maps, while effectively smoothing the data
over $\sim 125$ square-degrees, and still should provide a significant
test.  In order to use ENAs from the tail for identifying the
next-cloud, either pixels in the tail must be grouped to improve
statistics, or the comparison should wait for the better statistics
of future skymaps.

\begin{figure}[hb!]
\begin{center}
\plottwo{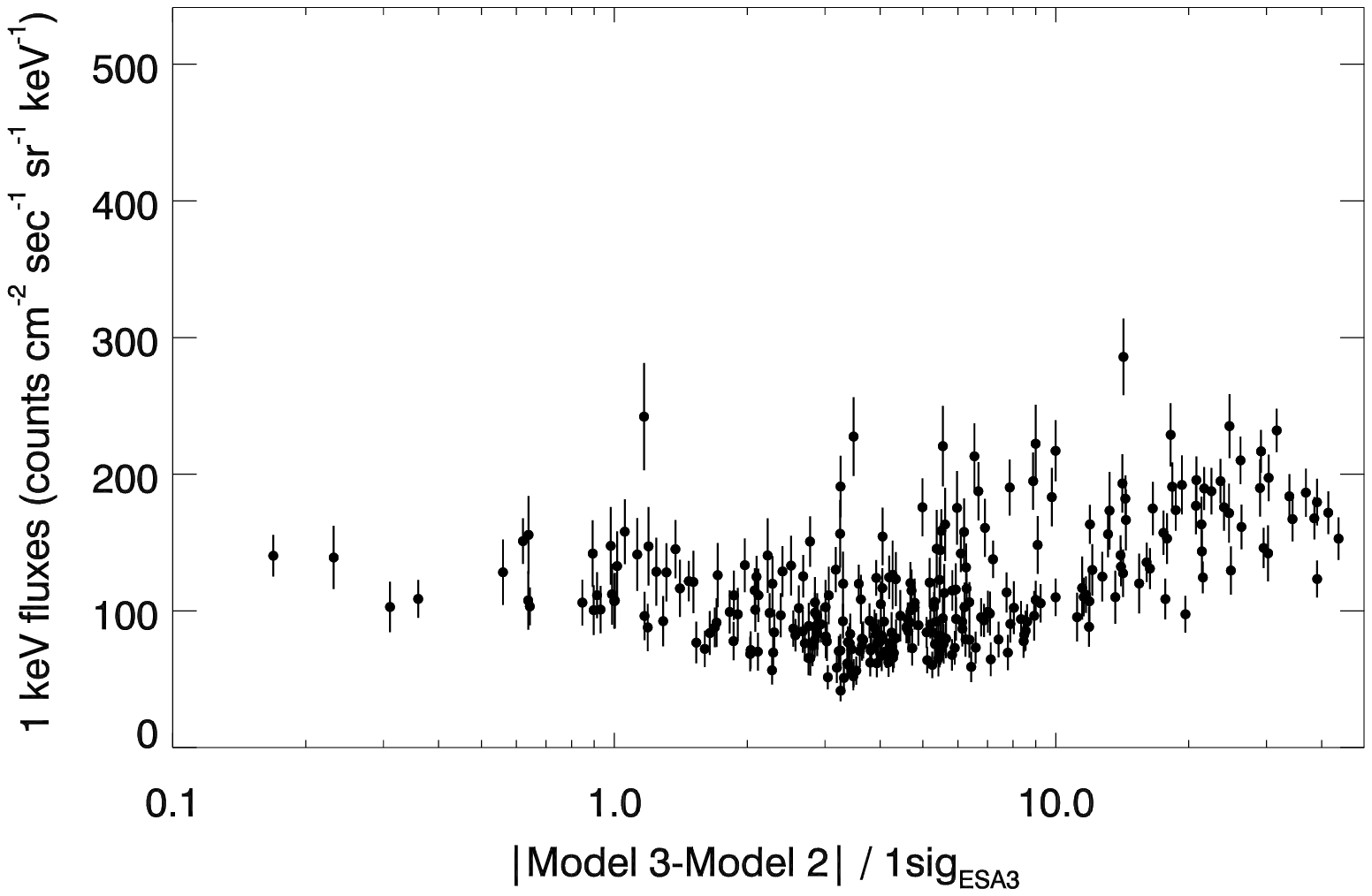}{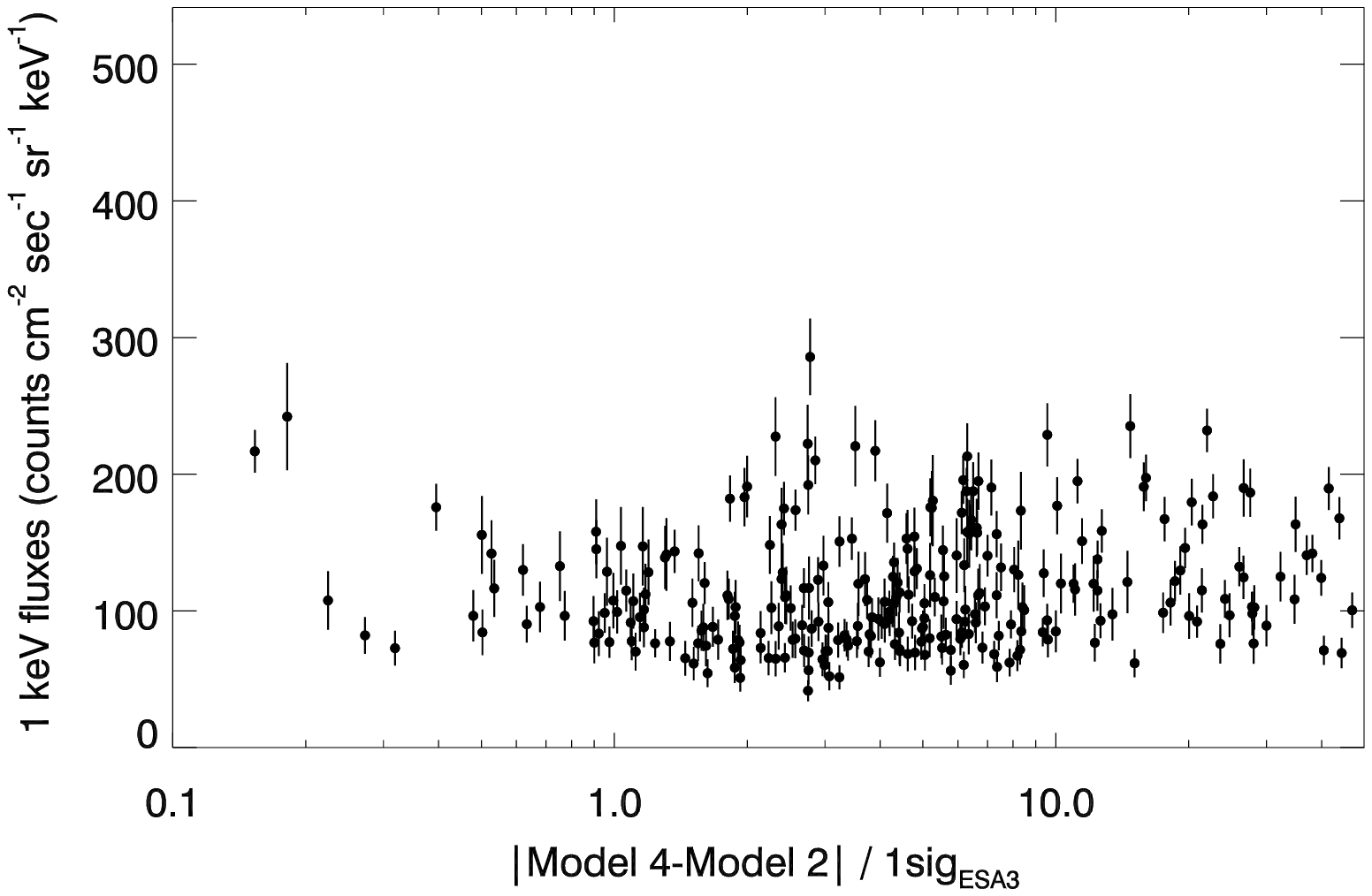}
\end{center}
\caption{The ESA 3 fluxes (ordinate) are compared to the significance
of the model differences
(abscissa) at each pixel on the sky for Models 3 and 2 (left)
and Models 4 and 2 (right).
The abscissa shows the differences in the fluxes of the two models, divided by
the $1\sigma$ measurement uncertainties on the ESA 3 data.  Only
"good" ESA 3 pixels with S/N$>3$ are plotted. The differences between
Models 3 and 2 are tested at the $3 \sigma$ level by 20\% of all ESA 3
pixels, while 49\% of all pixels test these differences at the
$1\sigma$ level, for example.  }
\label{fig:corr}
\end{figure}

The predicted ENA flux differences between the today-cloud and next cloud are
testable with IBEX data.  Twenty percent of the ESA 3 (1.1 keV) pixels with signal-to-noise S/N$> 3$
test the flux differences between Model 3 and Model 2 at the $3 \sigma$ level,
or \delmod/\dFesa3$>3$.  In addition, 49\% of the
pixels test these flux differences at the $1\sigma$ level, with
\delmod/\dFesa3$>1$ (Fig. \ref{fig:corr}).  Similar values are found
for comparisons between the predicted flux differences between Model 4
and Model 2, where 18\% of the pixels show model differences that are
larger than the $3 \sigma$ ESA3 flux uncertainties.

\begin{figure}[ht!]
\begin{center}
\includegraphics[width=0.70\textwidth]{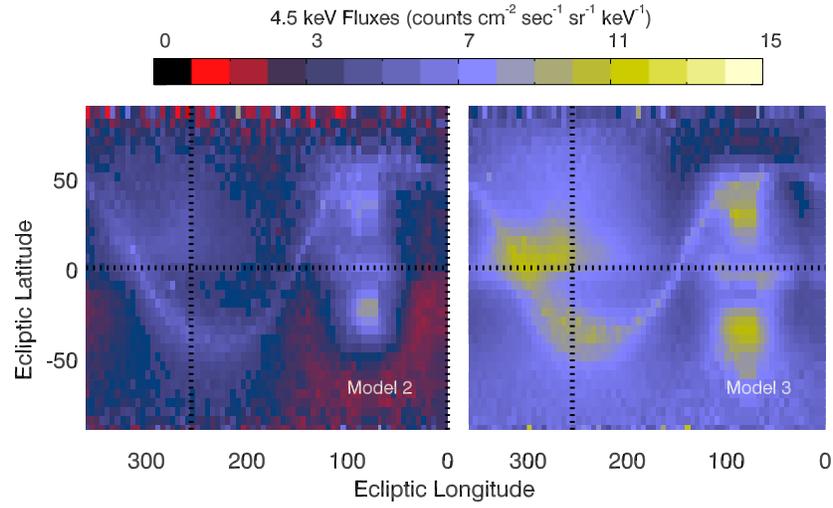}
\end{center}
\caption{The predicted ENA fluxes at 4.5 keV (ESA 6 energy passband)
are shown for Model 2 (left) and Model 3 (right).  }
\label{fig:4p5mult}
\end{figure}

\begin{figure}[hb!]
\plottwo{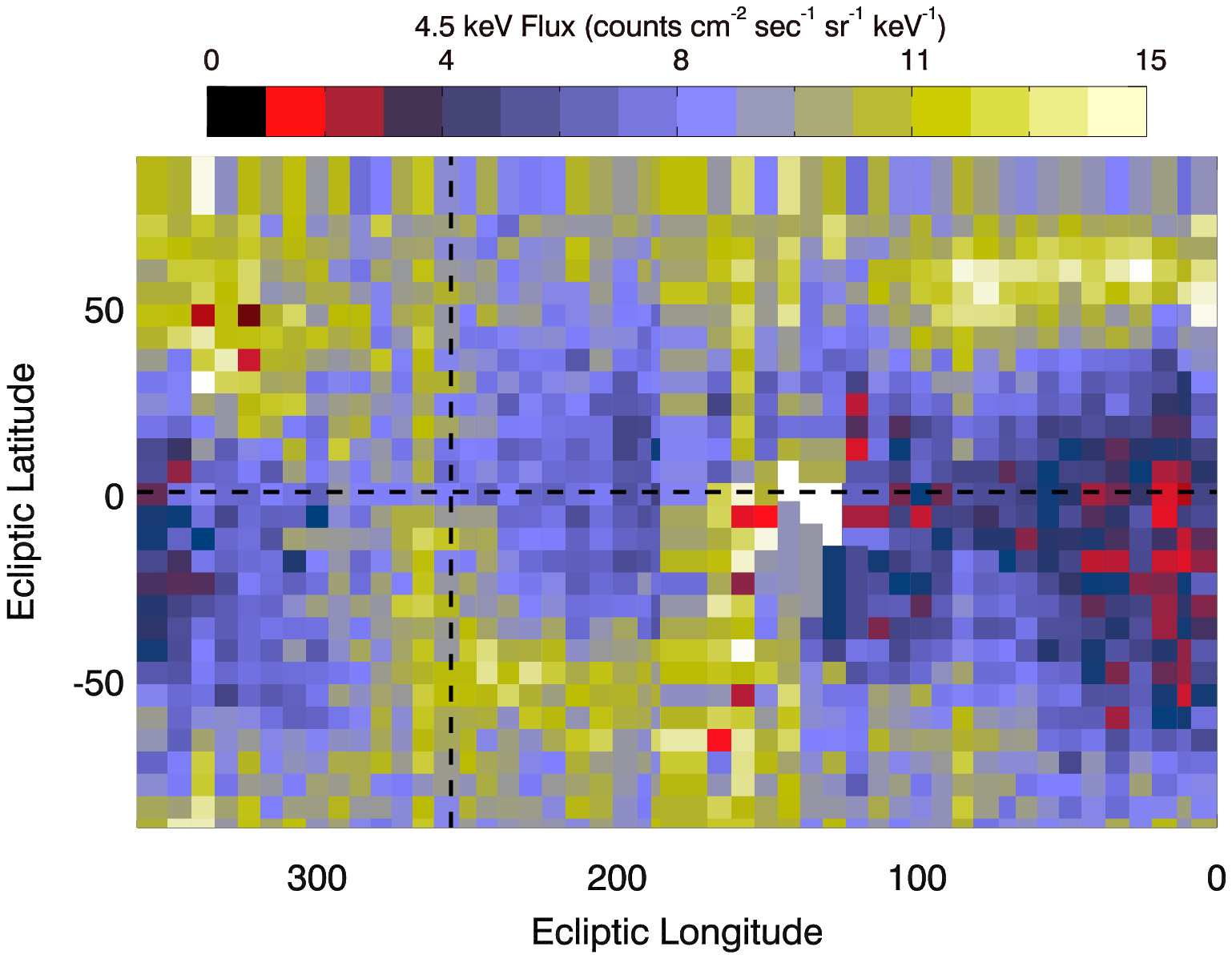}{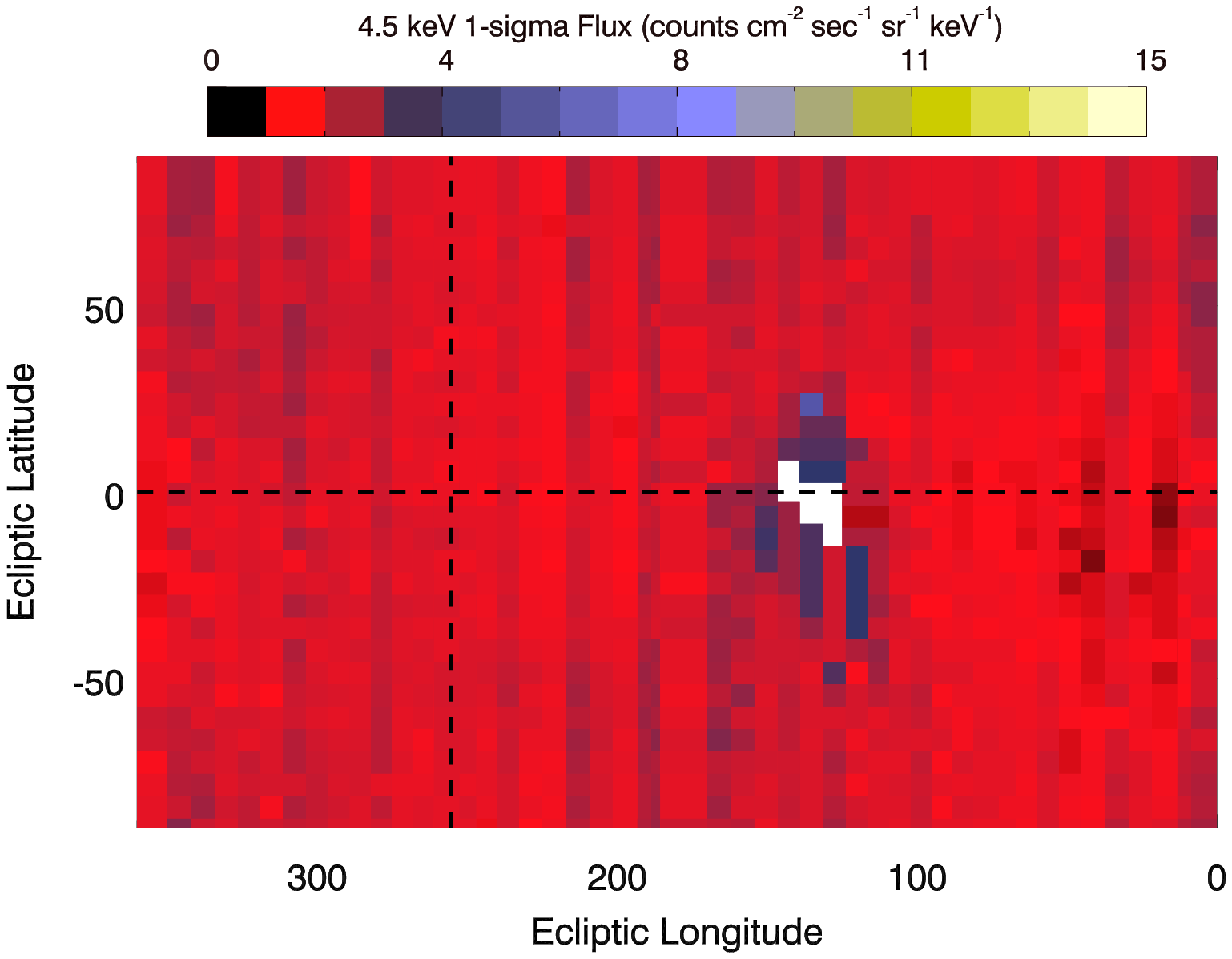}
\caption{The observed ESA 6 ENA fluxes at 4.5 keV are shown (left),
together with the $1 \sigma$ uncertainties on those fluxes (right).
}
\label{fig:4p5diff}
\end{figure}

We have also evaluated the variations in the 4.5 keV ENA fluxes for
the environment of the next cloud (Fig. \ref{fig:4p5mult}).  Although
the count rates in IBEX-HI ESA 6, at 4.5 keV, are lower by an order of
magnitude than at 1.1 keV (Fig. \ref{fig:4p5diff}, left), flux
variations are predicted to occur when the effect of the
increased interstellar density and velocity are included (Model 3 vrs.
Model 2), as well as when the ISMF direction varies (Model 4 vrs.
Model 2).  For example, Model 2 has fluxes 5--6 \counts\ at the
locations \elong,\elat=$253^\circ,-33^\circ$ and
\elong,\elat=$268^\circ,8^\circ$.  For the same locations, Model 3 has
fluxes a factor of $\sim 2$ higher.  Further study of the energy
spectrum of heliosheath ions is in progress, however, since the
relatively large tail brightness predicted at 4.5 keV by Model 2 is
difficult to distinguish in the data.

\section{Discussion }

If the Ribbon ENAs are produced as secondary ENAs beyond the
heliopause as suggested by McComas et al. (2009) and Schwadron et
al. (2009), and quantified by Heerikhuisen et al. (2010), then the
position and intensity of the Ribbon provides a robust diagnostic of
the interstellar magnetic field direction, neutral densities, and the
cloud ram pressure at the heliosphere.  Differences between the
velocity of ISM inside of the heliosphere and towards the nearest
stars in the upwind direction indicate that the Sun is near or at the
edge of an interstellar cloud.  Based on observations of the upwind
ISM, and assuming that the upwind cloud is in pressure equilibrium
with the heliospheric ISM, the next cloud is modeled with densities
that are $\sim50$\% larger, and a heliocentric velocity 9\% larger,
than the cloud today.  We predict the flux of ENAs from the next
interstellar cloud to surround the heliosphere, and compare those
predictions with the measurement uncertainties of the 1.1 keV ENA fluxes
detected by IBEX in its first 6 month skymap.  These results rely
explicitly on the assumption that the \citet{Heerikhuisen:2010ribbon}
model is a viable description of ENA production both for the cloud we
are in today and for the nearby cloud observed towards $\alpha$ Cen
and 36 Oph in the upwind direction.  Although our detailed conclusions
rely on the accuracy of the \citet{Heerikhuisen:2010ribbon} model,
this study is a useful gedanken experiment that will help us
understand the Ribbon sensitivity to variations in the properties of
the ISM around the heliosphere.

The variations in ENA fluxes predicted for entry into the next cloud
significantly exceed the measurement uncertainties for 20\% of the
ESA 3 pixels, which tend to be concentrated in the upwind hemisphere,
and the variations are larger by a factor of two in some regions.  The variations
occur because the relative contributions of magnetic pressure and
thermal ram pressure that deform the heliosphere are sensitive
functions of the boundary conditions imposed by the ISM.  If, in
addition, the direction of the interstellar magnetic field shifts by
as much as $28^ \circ$, which is slightly larger than the Ribbon
width, then significant differences in the ENA fluxes should be
observed in individual pixels near the Ribbon.  The heliosphere
regions with very low ENA fluxes, such as the tail, provide a test of
the cloud properties only if pixels are combined for better
statistical significance before comparison with model predictions.  The
long term capability of IBEX to realistically detect such variations,
which will also be superimposed on possible solar cycle variations,
requires that the efficiencies in the IBEX sensors (conversion,
scattering, sputtering in the conversion subsystem, secondary electron
emission at the detector foils, microchannel plate efficiencies, for
instance) either remain stable over years, or alternatively that the
instrument performances are tightly monitored.  IBEX-HI detector
sensitivity is continuously monitored in a number of ways,
such as comparison of coincident and non-coincident count rates 
\citep{FunstenHarperMcComas:2005}, and periodic gain tests.

Variations in the energy dependence and fluxes of ENAs will occur
because of the variation of solar wind properties over the solar
cycle.  These solar cycle contributions fortunately can be modeled in
detail using past and present data on the solar wind, and models of
the heliosphere response to these variations.  Every IBEX skymap is a
historical map of the solar cycle because of the energy dependence of
ENA travel times and cross sections (McComas et al. submitted, 2010),
so that unraveling the solar cycle dependence will simultaneously
constrain the ENA production models and improve future predictions of
the ENA variations due to the next interstellar cloud.

In this discussion we considered the scenario where the next cloud is
faster, slightly cooler, and more dense than the heliospheric ISM gas, as
expected from pressure equilibrium and observational data.  This study
is a proof-of-concept, since the properties of the cloud edges are not
established. The more extreme possibilities for the next galactic
environment to be encountered by the Sun include a hot plasma without
neutrals, and a cloud interface that is either evaporative or mixed
with hot plasma by shear flows. If interstellar clouds within 10 pc
are in pressure equilibrium, they will typically fill 20\% of the
sightline to the stars.  The intervening voids evidently will be
filled with the low density hot gas that creates the Local Bubble soft
X-ray emission, although the emissivity of this plasma is somewhat
uncertain because of solar wind contamination
\citep{Koutroumpaetal:2008}.  Should the heliosphere enter the diffuse
plasma attributed to the Local Bubble interior, both interstellar
neutrals and exo-heliospheric ENAs will vanish.  IBEX and other
spacecraft will readily detect this condition.  Another possibility
for the next solar environment would be an evaporative interface that
would form upwind between the heliospheric ISM and hot plasma.  Such
an interface will show steep increases in the cloud velocity and
pressure, and decreases in density, over spatial scales that are
determined by the angle between the ISMF and cloud surface
\citep{SlavinFrisch:2008}. 

The present study considers the ENA fluxes for two separate models of
the circumheliospheric cloud, but ignores possible variations due to
changes in the heliosphere configuration during the transitions
between the two clouds.  
The predicted thickness of the conductive boundary on the cloud around the heliosphere,
defined as where the temperature falls to 50\% of the asymptotic temperature, is
0.32--0.34 pc for an ISMF direction that makes an angle of
30\deeg\ with the cloud surface 
\citep[][Models 26 and 27, also see Fig. 2, where the cloud edge starts at 3 pc]{SlavinFrisch:2008}.  
In the upwind direction the outflow speeds in the conductive boundary are 20--30 \kms, and
opposite to the cloud motion, so that the Sun could
traverse the conductive boundary in approximately $\sim 12,000$ years for these models.  
Based on the above models, we expect the
change in heliosphere properties for such an environment to be clearly
observable in the resulting ENA flux detected by IBEX.  
A turbulent mixing layer will also produce strong gradients in the temperature and
ionization of the surrounding ISM \citep{Slavinetal:1993}.  
The ENA emissions for a conductive boundary on the surrounding cloud
are discussed in detail in \citet{GrzedzielskiBzowskietal:2010}, where
the Sun is estimated to emerge from the interface within $\sim 500$ years.
An alternative possibility is that the G-cloud may be denser than has been
assumed in this study.  If the interstellar \CaII\ absorption formed at the G-cloud velocity
is entirely within a few parsecs of the Sun, then comparisons between the
clouds in the $\alpha$ Cen and $\alpha$ Oph sightlines suggest a tiny cold cloud 
in addition to the warmer gas \citep{Frisch:2003apex}.

The comparisons in this paper are made without consideration of the
solar cycle, although the outer heliosheath regions respond to the
variations in the solar wind dynamic pressure and magnetic field that
characterize the solar activity cycle
\citep{WashimiTanaka:1999,SchererFahr:2003,ZankMueller:2003cycle,Pogorelovetal:2009solarcycle}.
Although the solar cycle will cause the heliosphere to expand and contract as
the solar wind dynamic pressure changes, these pulses travel
only a relatively short distance upstream of the heliopause ($\sim
100$ AU).  The influence of the solar cycle on ENA production and the
Ribbon phenomenon is not yet understood.  The Ribbon intensity may
vary over latitudes due to the ion energy differences and travel
times.  The extremely low levels of solar activity during the first
year of IBEX observations suggests the solar activity cycle variations
must first be understood before reaching a conclusion that we have
entered a new interstellar cloud
\citep{PogorelovOgino:2008,SternalScherer:2008ibexsolarcycle}.  As the
theoretical models of ENA production become increasingly robust, we
expect that studies such as this will yield definitive information on
both the heliosphere boundary conditions and the physical properties
of the interstellar cloud around the Sun.  Finally, while this study
has examined only one of the possible sources of the Ribbon currently
under discussion, the other ideas for producing the ribbon
\citep{McComas:2009sci} also generally invoke and seek to match up
with the orientation of the external IMF, so even if another
explanation eventually becomes accepted, it may still be possible to
directly detect the interstellar transition with IBEX.

\acknowledgments

We thank the IBEX team members.  This work was funded through the IBEX
mission, as a part of NASA's Explorer Program.  Jacob Heerikhuisen
would like to acknowledge support from the NASA IBEX program through
the grant NNX09AG63G.  We would like to thank Ed Roelof for his
help with the Compton-Getting corrections.




\clearpage
 \begin{table}
 \begin{center}
 \caption{Properties of Circumheliospheric Interstellar Material used for Models \tablenotemark{a} \label{tab:model}}
 \begin{tabular}{lccccc}
 \tableline\tableline
Quantity & LIC\tablenotemark{b} & Model 1\tablenotemark{c} & Model 2\tablenotemark{d} & Model 3\tablenotemark{e} & Model 4\tablenotemark{f}  \\ 
                & &  Original &  Today & Next (same ISMF) & Next (new ISMF)   \\ 
 \tableline
 \nHI\  (\cc)  & 0.19  & 0.15 & 0.15 &  0.23 &  0.23     \\
 \npro\  (\cc)  & 0.06 & 0.06 & 0.06 &  0.08 &  0.08  \\
 \vel\  (\kms)  & --26.3 & --26.4 & --26.4 & --28.8 & --28.8    \\
 \temp\  (K)  & 6300 & 6530 & 6530 & 5400  &  5400    \\
$|$\B  $| $ (\microG)& [2.7]\tablenotemark{g} & 3.  &  3.  &  2.8   &  2.8     \\
\B\ direction, \elong,\elat  &  & $237.^\circ,30 ^\circ$ &  $224^\circ,41^\circ$ &  $224^\circ,41 ^\circ$ &  $252^\circ,42 ^\circ$ \\
\B\ direction, \glong,\glat  &  & $22.^\circ,41 ^\circ$ &  $36^\circ,53^\circ$ &  $36^\circ,53 ^\circ$ &  $40.^\circ,32 ^\circ$ \\
Magnetosonic Mach &   1.0 &  1.1 &  1.1 &  0.8 & 0.8 \\
$~~~~$ number &&&&& \\
 \end{tabular}
 
 
\tablenotetext{a}{The 1 AU solar wind parameters used for all models are \npro=7.4 \cc, \temp=51,100 K, \vel=450 \kms, 
$|$\B $_\mathrm{radial} | = 37.5$ \microG.  The ISMF direction indicated by the center
of the Ribbon corresponds to $\lambda,~\beta = 221^\circ,~39^\circ$ (Funsten et al. 2009).}
\tablenotetext{b}{These values for the ISM forming the heliosphere boundary conditions
are based on Model 26 in \citet{SlavinFrisch:2008} and \citet{Witte:2004}.}
\tablenotetext{c}{\citet{Schwadron:2009sci} used this model \citep{Pogorelovetal:2008let} in the initial
analysis of IBEX data.}
\tablenotetext{d}{This model reproduces the IBEX ribbon \citep{Heerikhuisen:2010ribbon}.} 
\tablenotetext{e}{The next-cloud model, assuming the same ISMF direction as the today-model, Model 2.}
\tablenotetext{f}{The next-cloud model, assuming an ISMF direction that differs from Model 2. }
\tablenotetext{g}{Determined by assuming that thermal and magnetic
pressures are equal.}
 \end{center}
 \end{table}

\clearpage

\clearpage

\clearpage

\clearpage

\clearpage

\clearpage

\clearpage

\clearpage

\clearpage

\clearpage

\end{document}